\documentclass[12pt]{article}
\usepackage{mathrsfs}
\usepackage{amsthm}
\usepackage{natbib,graphicx,setspace,lscape,longtable,xr}
\usepackage{mathrsfs,amsmath,amssymb,color}
\usepackage{natbib,epsfig,pdfpages}
\usepackage{rotating}
\usepackage{float}
\usepackage{bm}
\usepackage[normalem]{ulem}
\usepackage{ragged2e}
\usepackage{enumerate}
\usepackage{float}    
\usepackage{subfig}
\usepackage{fancyhdr} 
\usepackage{lastpage}     
\graphicspath{{figures/}}
\usepackage{algorithm}
\usepackage{algorithmic}
\usepackage{bbm}
\usepackage{svg}
\usepackage{lipsum}
\usepackage{mathtools,cuted}
\usepackage{bbm}
\usepackage{tabularx}
\usepackage{booktabs}
\usepackage[symbol,multiple]{footmisc}
\usepackage{lineno,hyperref}
\usepackage{booktabs}
\usepackage{multirow}
\usepackage[toc,page]{appendix}
\usepackage{bigfoot}
\usepackage{makecell}

\DeclareNewFootnote{AAffil}[arabic]
\DeclareNewFootnote{ANote}[fnsymbol]

\usepackage{etoolbox}
\newcommand{\blind}{0}
\newtheorem{theorem}{Theorem}

\newtheorem{proposition}{Proposition}
\newtheorem{definition}{Definition}

\newtheorem{example}{Example}
\newtheorem{remark}{Remark}

\def\bO{\bm O}
\def\bn{\bm n}
\def\bLambda{\bm \Lambda}
\def\bTheta{\bm \Theta}
\def\bpi{\bm \pi}
\def\bb{\bm b}
\def\bz{\bm z}
\def\be{\bm e}
\def\one{{\bm 1}}
\def\bA{{\bm A}}
\def\bM{{\bm M}}
\def\bPsi{{\bm \Psi}}

\def\beq{\begin{equation}}
\def\eeq{\end{equation}}
\def\beqr{\begin{eqnarray}}
\def\eeqr{\end{eqnarray}}
\def\beqrs{\begin{eqnarray*}}
\def\eeqrs{\end{eqnarray*}}

\def\mN{\mathcal N}

\def\mR{\mathbb R}
\def\mbN{\mathbb{N}}

\def\mS{\mathcal S}

\def\bI{\mathbb I}

\def\mS{\mathcal S}

\def\bM{\bm M}

\def\mK{\mathcal K}

\def\mW{\mathcal{W}}

\def\argmax{\mbox{argmax}}
\def\diag{\mbox{diag}}

\newcommand{\balpha}{\boldsymbol{\alpha}}

\usepackage{hyperref}
 \hypersetup{
	colorlinks=true,
	filecolor=blue,
	citecolor = blue,
	urlcolor=cyan,
}

\modulolinenumbers[5]

\makeatletter
\patchcmd\maketitle{\def\@makefnmark{\rlap{\@textsuperscript{\normalfont\@thefnmark}}}}{}{}{}
\makeatother

\makeatletter
\def\thanksAAffil#1{
  \footnotemarkAAffil\protected@xdef\@thanks{\@thanks%
        \protect\footnotetextAAffil[\the \c@footnoteAAffil]{#1}}%
}
\def\thanksANote#1{%
  \footnotemarkANote%
  \protected@xdef\@thanks{\@thanks%
        \protect\footnotetextANote[\the \c@footnoteANote]{#1}}%
}
\makeatother

\DeclareFontFamily{U}{mathx}{}
\DeclareFontShape{U}{mathx}{m}{n}{<-> mathx10}{}
\DeclareSymbolFont{mathx}{U}{mathx}{m}{n}
\DeclareMathAccent{\widehat}{0}{mathx}{"70}
\DeclareMathAccent{\widecheck}{0}{mathx}{"71}

\begin{document}
\emergencystretch 3em

\def\spacingset#1{\renewcommand{\baselinestretch}%
{#1}\small\normalsize} \spacingset{1}

\date{}
\if0\blind
{
  \title{\bf Distributed Pseudo-Likelihood Method for Community Detection in Large-Scale Networks\thanksANote{This work was supported by the National Natural Science Foundation of China (grant numbers 12071477, 71873137, and 72271232) and the MOE Project of Key Research Institute of Humanities and Social Sciences (22JJD110001), and the fund for building world-class universities (disciplines) at Renmin University of China. The authors gratefully acknowledge the support of Public Computing Cloud, Renmin University of China.}}
\author{Jiayi Deng\thanksAAffil{Department of Statistics and Epidemiology, Graduate School of the PLA General Hospital, Beijing, China} , Danyang Huang\thanksAAffil{Center for Applied Statistics and School of Statistics, Renmin University of China, Beijing, China}$^{\ ,}$\thanksANote{Corresponding author, \href{mailto:dyhuang89@126.com}{\nolinkurl{dyhuang89@126.com}}} \ and Bo Zhang\footnotemarkAAffil[2]%
 }
  \maketitle
}\fi
\if1\blind
{
  \bigskip
  \bigskip
  \bigskip
  \begin{center}
    {\LARGE\bf Distributed Pseudo-Likelihood Method for Community Detection in Large-Scale Networks}
    \vspace{0.5em}\\  
\end{center}
  \medskip
} \fi

\bigskip
\begin{abstract}
This paper proposes a distributed pseudo-likelihood method (DPL) to conveniently identify the community structure of large-scale networks. Specifically, we first propose a {\it block-wise splitting} method to divide large-scale network data into several subnetworks and distribute them among multiple workers. For simplicity, we assume the classical stochastic block model. Then, the DPL algorithm is iteratively implemented for the distributed optimization of the sum of the local pseudo-likelihood functions. At each iteration, the worker updates its local community labels and communicates with the master. The master then broadcasts the combined estimator to each worker for the new iterative steps. Based on the distributed system, DPL significantly reduces the computational complexity of the traditional pseudo-likelihood method using a single machine. Furthermore, to ensure statistical accuracy, we theoretically discuss the requirements of the worker sample size. Moreover, we extend the DPL method to estimate degree-corrected stochastic block models. The superior performance of the proposed distributed algorithm is demonstrated through extensive numerical studies and real data analysis.
\end{abstract}

\noindent%
\vfill

\newpage
\spacingset{1.75} 
\section{Introduction}
\label{sec:intro}

Network community detection aims to cluster network nodes into different groups such that the connectivity intensity is higher within a group than between groups \citep{girvan2002community, newman2004finding}. This problem is a fundamental issue in network analysis, with wide applications in computer science \citep{agarwal2005beyond}, social science \citep{zhao2011community}, and biology \citep{nepusz2012detecting}. With the rapid development of information technology, we often encounter large-scale network data; however, the entire collected dataset cannot always be distributed in a single machine. This can be attributed to the following reasons. First, the datasets of various applications are significantly large to be stored and examined conveniently on a single machine. Second, privacy considerations may make it difficult or even impossible to pool separate datasets into a single machine. Therefore, community detection algorithms should be designed for network data stored on many connected machines, referred to as distributed systems.

Using a distributed system, large-scale data can be divided into many small subsamples. Subsequently, the computational problem can be decomposed and solved in a parallel manner. The final estimate can be reasonably obtained using the estimates or intermediate outputs. Here, we consider a ``master-and-worker'' distributed system, which comprises many {\it workers} and a {\it master}. A worker is a local machine with reasonable storage and computing power for storing the subsample and performing calculations based on the subsample. The master is a central computer responsible for collecting and aggregating intermediate results from different workers based on the subsamples. Information transfer between different computers is referred to as {\it communication}. The master-and-worker distributed system is based on the assumption that communication is only allowed between the master and worker.

Communication cost is defined as the total number of bits exchanged between the workers and the master, which can be expensive in distributed systems owing to the limited bandwidth \citep{zhang2013communication, shamir2014communication, garg2014communication}. Therefore, several studies have focused on designing communication-efficient distributed algorithms. For instance, the one-shot approach requires only one round of communication between the master and each worker \citep{zhang2013communication, lee2015communication, battey2018distributed, fan2019distributed}. Communication-efficient multi-round methods conduct multiple rounds of communication between the workers and the master to refine the estimation efficiency \citep{shamir2014communication, wang2017efficient, wang2018giant, jordan2019communication, chen2021first}. Notably, a distributed community detection framework in network data is extremely different from the existing distributed settings.

\begin{figure}[!htb]
 \centering
 \includegraphics[width=0.65\linewidth]{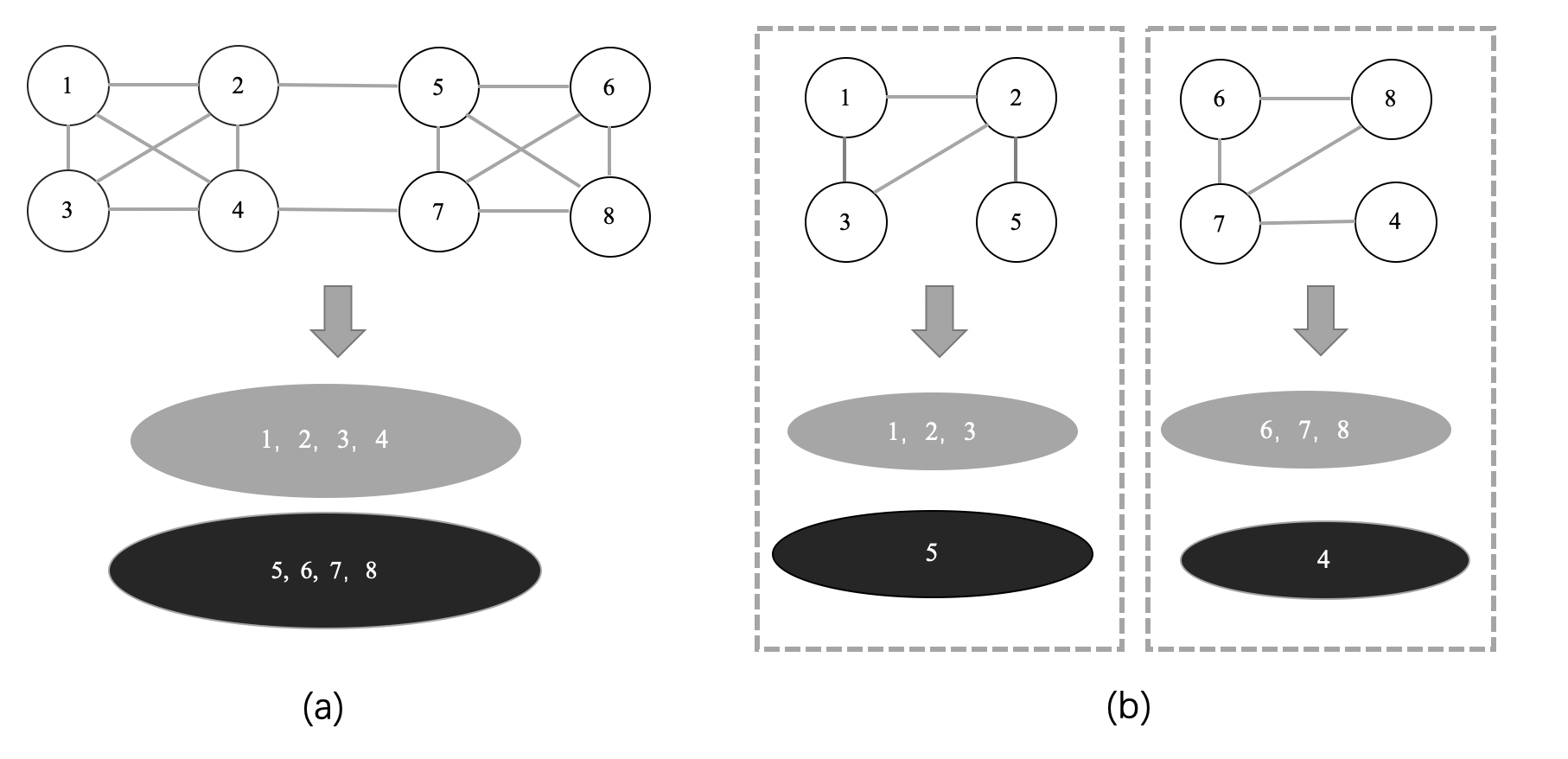}
 \caption{A network with eight nodes and two communities. (a) Community detection for the entire network; the communities are represented by gray and black circles. (b) Community detection for subnetworks in two workers. The local label estimates from each worker need to be aligned to match the community labels estimated from the entire network.}\label{fig: community}
\end{figure}

Specifically, there are two main challenges to community detection based on the distributed system. The first challenge lies in effectively partitioning interdependent network data into multiple workers. Unlike independent observations, network nodes are intricately connected. Therefore, a naive splitting based on individual nodes may result in the loss of critical connection information, leading to a considerable deterioration in the result of community detection. The second challenge is matching the label estimates of multiple workers. Consider a master-and-worker distributed system wherein each worker estimates the community labels of its local subsample and then the master combines the findings to obtain the community labels of the entire network. For example, in Figure \ref{fig: community}, the left panel illustrates the community detection results for an entire network, while the right panel displays the community labels estimated from each subnetwork. In the global estimator, nodes \{1,2,3,4\} are assigned to cluster 1 in gray. While in the first local estimator, nodes \{1,2,3\} are assigned to cluster 1 in gray, and in the second local estimator, node 4 is assigned to cluster 2 in black. Therefore, the local label estimates must be aligned to match the global estimator. For a subnetwork with $K$ communities, various potential label assignments exist (i.e., $K!$). The alignment of the local label estimates of all subnetworks is a complex problem \citep{yang2015divide, mukherjee2021two}.

In recent years, community detection based on distributed systems has been increasingly discussed \citep{chen2010parallel, sun2019distributed, mukherjee2021two, wu2023distributed}. Some methods use each worker to estimate the local community labels based on their local subnetwork, where the subnetwork is induced only by connections within the subset of nodes stored in a worker \citep{yang2015divide, mukherjee2021two}. Such methods require the alignment of local label assignments. Furthermore, approaches based on the spectral sparsification technique have also been developed \citep{chen2016communication, sun2019distributed}. In particular, each worker sends a sparsified local subnetwork to the master. Thereafter, the master constructs a sparse network, and the existing community detection method can be adopted in this network to obtain the community labels of all nodes. Although these distributed community detection methods have computational advantages over traditional community detection algorithms, they use information from the subnetwork instead of the entire network, resulting in inevitable information loss. Because observed connections are critical information in network community detection, we aim to develop a distributed approach to identify community structures by fully utilizing all edges from the sample.

In this paper, we propose a novel distributed pseudo-likelihood method (DPL) for community detection in large-scale networks. To divide a large-scale network, we first develop the {\it block-wise splitting} method based on its corresponding adjacency matrix. This approach ensures that all observed connections are distributed across workers. Under the stochastic block model (SBM) proposed by \cite{hoeffding1963probability}, each worker focuses on identifying the community labels of {\it in-worker nodes} by optimizing the local pseudo-likelihood. Each worker then broadcasts a local label estimator to the master, only considering the communication cost of the order $O(N/R)$ bits, where $R$ is the number of workers. As for the master, the label estimator of all nodes can be updated simply by combining the local estimators from the workers without requiring alignment. The updated result in the master can then be broadcasted to the workers, which is an initial estimator of each worker for the next iteration. This procedure requires only $O(NR)$ bits for communication. Consequently, the DPL framework can be easily applied across multiple iterations.

The novelty of this work can be summarized as follows: (1) \textbf{Computational efficiency}: the DPL method is computationally efficient with a complexity of $O(Nn\rho_N)$, as demonstrated in Proposition \ref{pro: computational}. This complexity is notably lower than that of existing methods. The proposed method enables multiple workers to share computational tasks for large-scale networks and can effectively update global estimates by combining local estimates without the complex process of aligning assignments. (2) \textbf{Storage efficiency}: the proposed \textit{block-wise splitting method} ensures that the distributed system records all connection information and prevents duplication of adjacency matrix storage across different workers. In contrast, existing distributed community detection methods, such as those by \cite{yang2015divide, mukherjee2021two}, and \cite{wu2023distributed}, rely on repeated storage of certain entries in adjacency matrices for label alignment.

The remainder of this paper is organized as follows. In Section 2, we briefly review related work. In Section 3, we introduce a distributed pseudo-likelihood method for community detection in large-scale networks. In Section 4, the experimental results of the proposed method are presented, and further discussion is provided in Section 5. All proofs are presented in the Appendix.

\section{Related Work}

Distributed statistical inference has drawn increasing attention for solving supervised learning problems with independent samples in various scenarios, including generalized linear models \citep{chen2014split, battey2018distributed, zhu2021least}, quantile regression models \citep{volgushev2019distributed, chen2020distributed}, principal component analysis \citep{garber2017communication, fan2019distributed}, and high-dimensional M-estimators \citep{shamir2014communication, jordan2019communication, fan2021communication}. Most of these are multi-round approaches that communicate multiple rounds between workers and the master to refine the estimation efficiency. As \cite{fan2021communication} pointed out, multi-round methods can achieve optimal statistical precision under broader settings than the one-shot approach \citep{zhang2013communication, lee2015communication}.

However, the aforementioned studies cannot be directly adopted to solve community detection for large-scale networks, which is an unsupervised learning task with dependent network nodes. We intend to address this problem. Moreover, for multi-round approaches, the communication cost of distributed statistical inference requires at least $O(dR)$ bits in each iteration, where $d$ is the dimension of the parameter. The parameter of interest for this study is the community labels of all nodes, which are $N$-dimensional. This notion implies that the proposed DPL method has a communication efficiency of the same order as that of the existing distributed statistical inference methods for independent samples.

Several community detection methods have been proposed, and they can be roughly categorized into two types. The first type comprises optimization-based algorithms that are derived independently of any specific model assumptions. These approaches typically involve addressing a global optimization problem, such as normalized cuts \citep{shi2000normalized}, modularity \citep{newman2004finding}, and multiple objectives \citep{liu2014multiobjective, pizzuti2017evolutionary}. Majority of related methodologies include spectral clustering algorithms \citep{ng2001spectral, von2007tutorial, li2022divide}, modularity-based algorithms \citep{newman2004finding, zeng2018scalable}, and evolutionary computation-based (EC-based) algorithms \citep{pizzuti2017evolutionary, zhang2018network, li2020multi, su2021parallel,yin2021multi}. Despite their advantages of being model-free and adaptable to complex networks, these algorithms face challenges in discussing the consistency of clustering results without explicit model assumptions.

The second type consists of probabilistic graphical model-based algorithms, which are developed based on specific model assumptions. Among these, the stochastic block model (SBM, \citealt{holland1983stochastic}) stands as one of the fundamental models for networks with community structures. Numerous extensions of SBM exist, such as degree-corrected SBM (DCSBM, \citealt{karrer2011stochastic}), mixed membership SBM (MMSBM, \citealt{airoldi2008mixed}), dynamic SBM (DynSBM, \citealt{matias2017statistical}), among others. Previous approaches for estimating SBM and its variants primarily include spectral clustering \citep{rohe2011spectral, lei2015consistency}, semidefinite programming-based methods \citep{chen2014clustering, cai2015robust}, pseudo-likelihood methods \citep{amini2013pseudo, wang2021fast}, Bayesian approaches \citep{yang2015bayesian, yang2017stochastic}, and hierarchical clustering \citep{lyzinski2016community, li2022hierarchical}. These model-based algorithms have well-founded theoretical guarantees for their clustering results and provide meaningful statistical insights into network structures. However, they are computationally expensive when dealing with large-scale networks. To enhance efficiency, community detection algorithms for large-scale SBMs adopt techniques like network subsampling \citep{deng2021subsampling, zhang2022randomized} and distributed computing methods \citep{zhang2022distributed, wu2023distributed}.


\begin{table}[ht]
\caption{Comparison of different distributed/parallel community detection algorithms. The prototype algorithms include spectral clustering (SC, \citealt{rohe2011spectral}), semidefinite programming-based methods (SDP, \citealt{chen2014clustering, cai2015robust}), and pseudo-likelihood methods (PL, \citealt{amini2013pseudo}). The computational complexity is provided for a network comprising $N$ nodes, with its network density denoted by $\rho_N \in (0,1)$.}\label{tab: compare}
\resizebox{\columnwidth}{!}{
\centering
\begin{tabular}{cccccc}
  \hline
 \makecell[c]{Distributed/parallel\\ algorithm} & \makecell[c]{SBM-based \\ algorithm} & \makecell[c]{ Network \\ information}  & \makecell[c]{Computational \\ complexity}  & \makecell[c]{Prototype \\ algorithms} & \makecell[c]{Network distributed \\ storage}\\
  \hline
 \makecell[c]{MsgPassing \\ \cite{chen2016communication}}  &No & Subnetwork & $O(Nn)$  & SC  & Yes \\
  &  &  &  &  \\
 \makecell[c]{Blackboard \\ \cite{chen2016communication}} &No & Subnetwork  & $O(Nn)$  &  SC & Yes \\
  &  &  &  &  \\
   \makecell[c]{ DC \\ \cite{yang2015divide}} &No & \makecell[c]{Entire \\ network}  & $O(N^{2})$  & SDP & No \\
    &  &  &  &  \\
 \makecell[c]{PMOEA \\ \cite{su2021parallel}} &No & \makecell[c]{Entire \\ network} & $O(N^2\rho_N)$ &SC & No\\
     &  &  &  &  \\
 \makecell[c]{PSC \\ \cite{chen2010parallel}} &No & \makecell[c]{Entire \\ network}  & $O(Nn)$ & SC& Yes\\
     &  &  &  &  \\
 \makecell[c]{ PACE \\ \cite{mukherjee2021two}} &Yes & Subnetwork & $O(N^{2})$  &  \makecell[c]{SC, SDP,\\ PL, etc.} & Yes \\
      &  &  &  &  \\
 \makecell[c]{ GALE \\ \cite{mukherjee2021two}} &Yes  &Subnetwork  & $O(N^2)$ &  \makecell[c]{SC, SDP, \\ PL, etc.} & No \\
 &  &  &  &  \\
  \makecell[c]{DCD \\ \cite{wu2023distributed}} &Yes  &Subnetwork  & $O(N^2)$  &  \makecell[c]{SC} &Yes \\
 &  &  &  &  \\
 DPL & Yes & \makecell[c]{Entire \\ network}   & $O(Nn\rho_N)$  &PL &Yes \\
   \hline
\end{tabular}
}
\end{table}

To clarify our contributions, we conducted a comparative analysis of the proposed method against existing distributed/parallel community detection approaches, as detailed in Table \ref{tab: compare}. First, in scenarios where the network density $\rho_N \to 0$ and the worker sample size $n << N$, the proposed DPL method demonstrates superior computational efficiency compared to the examined algorithms. This advantage is theoretically supported by Proposition \ref{pro: computational}. Second, unlike the methods introduced by \cite{chen2016communication} and \cite{mukherjee2021two}, the DPL method makes use of entire network information rather than relying on partial edges to estimate community labels. Third, compared to the approaches developed by \cite{yang2015divide} and \cite{mukherjee2021two}, DPL avoids the complex process of local estimate alignment. In comparison to the EC-based algorithm by \cite{su2021parallel}, DPL operates in a distributed system where each worker updates local labels using only relevant connection information. Conversely, the EC-based algorithm employs the entire network to identify each local community structure. This makes the proposed method more computationally efficient.

In Section 4, we will perform a series of simulation studies and conduct real data analyses to compare the efficiency and accuracy of the proposed DPL method with the spectral clustering-based method \citep{chen2010parallel}, the EC-based algorithm \citep{su2021parallel}, and the distributed algorithms to estimate SBM \citep{mukherjee2021two, wu2023distributed}.

\section{Distributed Pseudo-Likelihood Method for Community Detection}

In this section, we briefly review the stochastic block model and degree-corrected SBM (DCSBM). Under a distributed system, we propose a block-wise splitting method to divide the entire network into several subnetworks to be stored by different workers. Under SBM, we provide an estimation method for each worker based on its local subnetwork. Subsequently, we develop a distributed network community detection method to estimate the SBM. Moreover, we provide a theoretical discussion on the subsample size of each worker, computational complexity, and communication cost of the proposed method. Finally, we extend the DPL method to degree-corrected SBM.

\subsection{Stochastic Block Model and Degree-Corrected SBM}

Consider an undirected network with $N$ nodes, which can be clustered into $K$ groups. Let vector $\bz$ denote the true community label, where $\bz \in [K]^N$ and $[K]=\{1,\cdots, K\}$. Furthermore, we define a symmetric matrix, $\bTheta=(\theta_{kl}) \in (0,1)^{K\times K}$, where $\theta_{kl}$ represents the connectivity probability between communities $k$ and $l$. Consider an $N\times N$ symmetric matrix $\bA=(a_{ij}) \in \{0,1\}^{N \times N}$ as an adjacency matrix. That is, if there is an edge between node pairs $(i,j)$, then $a_{ij}=1$; otherwise, $a_{ij}=0$ for all $1\le i\neq j\le N$, and $a_{ii}=0$ for $i=1,\cdots, N$. Then, we can define the SBM as follows:
\begin{definition}[Stochastic block model]
Suppose the latent label variables $\bz=(z_1, \cdots, z_{N})$ are drawn independently from ${\rm Multinomial}(\bpi)$, where $\bpi=(\pi_1,\cdots, \pi_K)$ and $\sum_{k=1}^K\pi_k=1.$ Furthermore, conditional on the community labels, the observed edges $a_{ij}$ ($i < j$) are independent Bernoulli variables with $P(a_{ij}=1|\bz) =\theta_{z_iz_j}$.
\end{definition}

In real-world networks, there are a few ``hub'' nodes with many connections, whereas most nodes have few connections. To incorporate degree heterogeneity within communities, \cite{karrer2011stochastic} proposed the degree-corrected stochastic block model (DCSBM) as an extension of SBM. A detailed definition of DCSBM is provided below.
\begin{definition}[Degree-corrected SBM] Let $\alpha_{i}$ represent the degree heterogeneity parameter of node $i$, and $\balpha= (\alpha_1,\cdots, \alpha_N)$. For any $i<j$, we assume the edge variables $a_{ij}$ are mutually independent Poisson variables with $E(a_{ij}|\bz)=\alpha_i\theta_{z_i z_j} \alpha_j$. Furthermore, the condition $\sum_{i}\alpha_i/N=1$ is added to ensure model identifiability \citep{zhao2012consistency}.
\end{definition}

\begin{table}[ht]
\centering
\caption{Notations.}\label{tab: notations}
\begin{tabular}{lll}
  \hline
  notations & & definitions \\
  \hline
  $N$ & & number of nodes in entire network\\
  $K$ & & number of communities in entire network \\
$\bz \in [K]^N$ & & the true community label \\
$\bTheta = (\theta_{kl}) \in (0,1)^{K\times K}$ & & connectivity probability matrix\\
$\bA \in \{0,1\}^{N\times N}$ & & adjacency matrix of entire network\\
 $\bpi \in (0,1)^{K}$ & &  probability vector of the node assignment \\
 $\balpha \in \mR^{N}$ & & degree heterogeneity parameter\\
 $R$ & & number of workers \\
 $\mN=[N]$ & &entire node set \\
 $\mN_r \subset [N]$ & & in-worker nodes \\
 $\bA_r \in \{0,1\}^{n\times N}$ & & sub-adjacency matrix\\
$\be \in [K]^N$ & &an initial label vector\\
$\bLambda \in \mbN^{K\times K}$ & & Poisson mean matrix\\
\hline
\end{tabular}
\end{table}

The notations used in this work are summarized in Table \ref{tab: notations}. Note that the problem of community detection is to infer the unknown community labels $\bz$ from the observed adjacency matrix $\bA$. In this work, we investigate identifying community labels for assortative networks \citep{amini2018semidefinite}. In other words, we assume $\max_{k\neq l} \theta_{kl}< \min_{k}\theta_{kk}$. Subsequently, we develop distributed community detection methods for SBM and DCSBM.

\subsection{Block-Wise Splitting}

\begin{figure}[!h]
 \centering
 \includegraphics[width=0.65\linewidth]{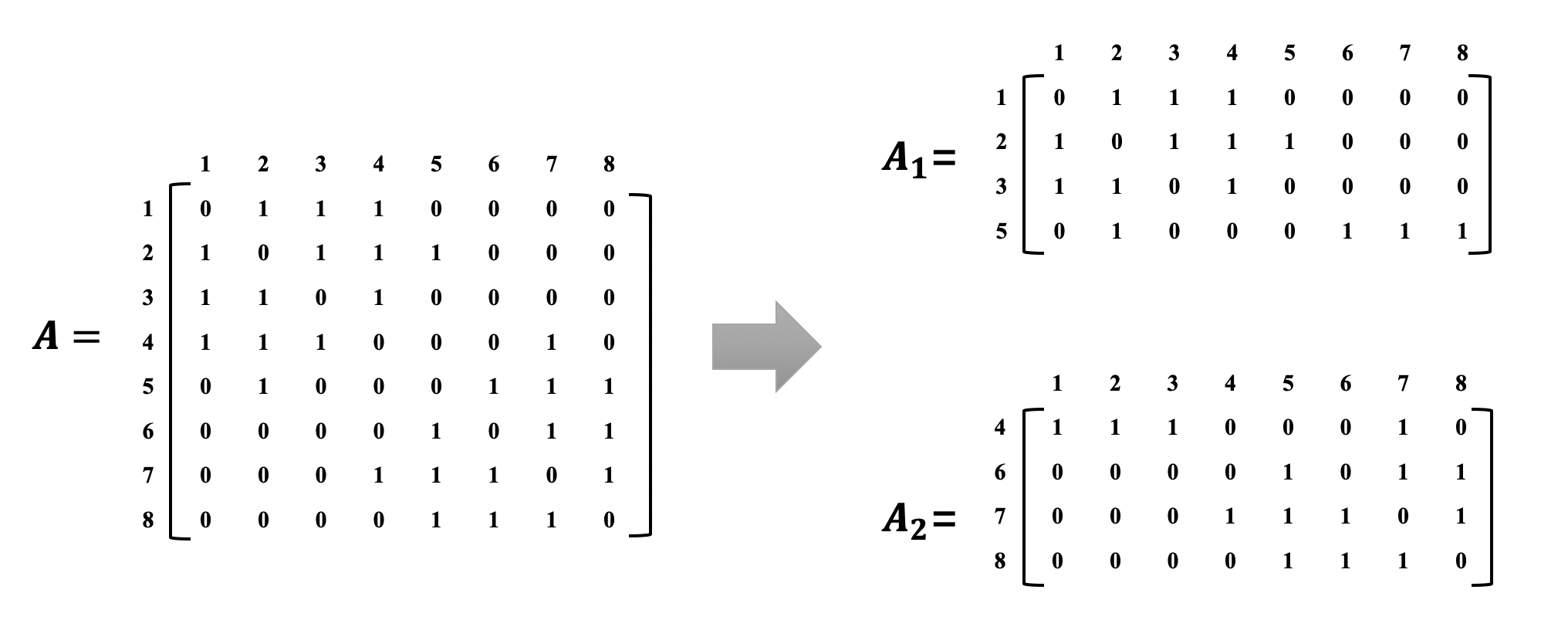}
 \caption{An example of the block-wise splitting method. The left panel shows the adjacency matrix of an entire network with eight nodes, where the nodes are uniformly divided into two blocks. For example, the 1st block contains four nodes, namely \{1,2,3,5\}. By block-wise splitting the adjacency matrix, we obtain two sub-adjacency matrices shown in the right panel, namely $\bA_1$ and $\bA_2$. Then, the subnetworks induced by two sub-adjacency matrices are distributed among the workers.
 }\label{fig: block-wise}
\end{figure}

We propose a {\it block-wise splitting} method to divide a large-scale network into many subnetworks for distribution among multiple workers. Let $\mN=\{1,\cdots, N\}$ represent the full node set that can be uniformly and randomly divided into $R$ disjoint subsets. There are subsets of nodes $\{\mN_r\}_{r=1}^{R}$ such that $\bigcup_{r=1}^{R}\mN_r=\mN$ and $\mN_{r_1}\bigcap\mN_{r_2}=\emptyset$ for $r_1\neq r_2.$ For simplicity, we assume a worker sample size of $|\mN_r|=N/R=n$, where $N/R$ is an integer. Then, the block-wise splitting method is provided as follows:

\begin{definition}[Block-wise splitting]\label{def: block} Define $n\times N$ dimension sub-adjacency matrices as $\bA_r=(a_{ij})_{i\in \mN_r, j \in \mN}$, for $r=1,\cdots, R.$ Thereafter, the adjacency matrix $\bA$ can be block-wise divided by $\bA=(\bA_r)_{r=1}^{R}$. We assign the subnetwork induced by $\bA_{r}$ to worker $\mW_r$ for $r=1,\cdots, R.$ Accordingly, the entire network is block-wise distributed among workers $\mW_1,\cdots, \mW_R$.
\end{definition}
\noindent Notably, the subnetwork recorded by each worker is formed by those connectivities related to its corresponding node set $\mN_r$. For convenience, we refer to the nodes in $\mN_r$ as {\it in-worker nodes} for the $r$th worker and the other nodes in the subnetwork as {\it the related nodes}. To illustrate the block-wise splitting method, an example is shown in Figure \ref{fig: block-wise}.

The block-wise splitting approach has the following appealing features. First, the subnetwork stored by each worker can be regarded as a subsample of the entire network. This method is easy to implement in practical applications. Second, the block-wise splitting method ensures storage efficiency by avoiding duplication of adjacency matrix information across different workers. This approach maintains the aggregated storage cost of sub-adjacency matrices equivalent to that of the entire adjacency matrix, effectively preventing an increase in overall storage demands. Third, this distributed method fully preserves the connectivity information between each in-worker node set and all network nodes, thereby ensuring the storage of all information from the entire network. Moreover, this feature helps match the label assignments across different workers. Lastly, based on the distributed samples, we introduce the parameter estimation procedure for SBM on each worker.

\subsection{Parameter Estimation on Workers}

We first introduce count statistics to describe the connectivity distribution of in-worker nodes. According to \cite{amini2013pseudo}, in order to simplify the likelihood function, we relax the symmetric constraints in the sub-adjacency matrix denoted by $\bA_r$. Consequently, we can treat rows of $\bA_r$ as independent variables. Denote the sub-adjacency matrix by $\bA_r=(a_{r,i'j})$, where $i'$ represents the node index of the worker, and $1\le i' \le n$. Next, based on the label vector $\bz$, for any $i'$th row, a count statistics $\bb_{r,i'}(\bz)$ is defined as $\bb_{r,i'}(\bz)=\{b_{r,i'k}(\bz)\}_{k=1}^{K}$, where $b_{r,i'k}(\bz)$ is given as follows:
\beq \label{eq: count}
b_{r,i'k}(\bz)= \sum_{j=1}^{N}a_{r,i'j}\bI(\bz_j=k),
\eeq
where $\bI(\cdot)$ is an indicator function. In other words, $b_{r,i'k}(\bz)$ represents the number of connectivities between node $i'$ and the nodes in the $k$th cluster. Since $b_{r,i'k}(\bz)$ is the sum of Bernoulli variables, it can be treated as a Poisson variable. However, as the label vector $\bz$ is a latent variable, the count statistics $\bb_{r,i'}(\bz)$ are also unobservable. To address this issue, we introduce an initial label vector $\be=(e_1,\cdots, e_N) \in [K]^N$ to replace $\bz$; therefore, the corresponding count statistics become $\bb_{r,i'}(\be)$. For convenience, we refer to $\bb_{r,i'}(\be)$ as $\bb_{r,i'}$.

We then provide the pseudo log-likelihood associated with $\{\bb_{r,i'}\}_{i'=1}^{n}$ and the parameters $(\bpi, \bTheta)$. Let $\bz_{r}=(z_{i})_{i\in \mN_r}$ denote the true labels of in-worker nodes. Then, for each node $i'$, conditional on the true label $\bz_r$ with $z_{r,i'}=l$, $b_{r,i'k}$ is considered a Poisson variable with strength parameter $\lambda_{lk}$. Furthermore, let $\Lambda= (\lambda_{lk})\in\mR^{K\times K}$ and $\lambda_{l}=\sum_{k}\lambda_{lk}$. Then, under the SBM framework, the pseudo log-likelihood function can be given as follows:
\beq\label{eq: SBM-likelihood}
\ell_{\rm SBM}(\bpi, \bLambda;\{\bb_{r,i'}\}_{i'=1}^{n})= \sum_{i'=1}^{n} \log\Big\{\sum_{l=1}^{K}\pi_{l}\exp{(-\lambda_{l})} \prod_{k=1}^{K} (\lambda_{lk})^{b_{r,i'k}}\Big\}.
\eeq

Furthermore, we discuss the parameter estimation. Let $\hat{\be} \in [K]^N$ be the initial label estimator of all network nodes. For each worker $\mW_r$, let $\hat{\be}_r=(\hat{\be}_i)_{i\in \mN_r}$ denote the initial label of the in-worker nodes. Next, we adopt an iterative algorithm to update the local label estimator $\hat{\be}_{r}$ and the parameter estimates $(\hat{\bpi}, \hat{\bLambda})$. Specifically, we first update the parameter estimates $(\hat{\bpi}, \hat{\bLambda})$ by using the expectation maximization (EM) algorithm to maximize \eqref{eq: SBM-likelihood}. Second, given the estimated $(\hat{\bpi}, \hat{\bLambda})$, we update $\hat{\be}_{r}$ as the most likely label for each node based on the EM convergence results. We discuss the iterative estimation algorithm in detail as follows.

\begin{figure}[!htb]
\centering
\includegraphics[width=0.85\linewidth]{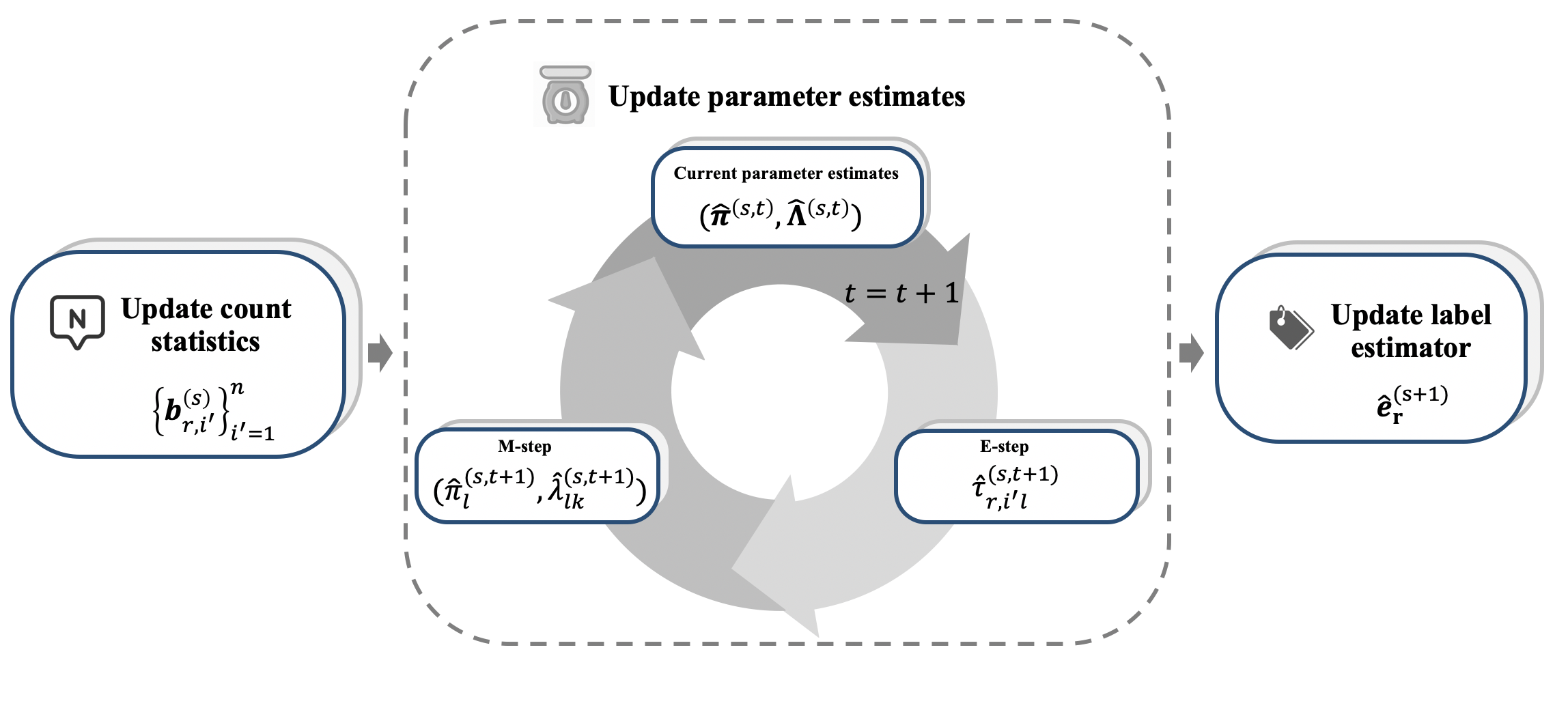}
 \caption{Procedures for updating the in-worker node label estimator in the $(s+1)$th iteration are determined mainly by the EM algorithm. This algorithm is used to update the parameter estimates $(\hat{\bpi}, \hat{\bLambda})$ and $\hat{\tau}_{r,i'l}$ ($1\le i'\le n, 1\le l \le K$), based on the local subnetwork.}\label{fig: worker-estimation}
\end{figure}

\begin{itemize}
\item {\bf Step 1} (Update count statistics). In the $(s+1)$th iteration, given the initial labeling $\hat{\be}^{(s)}$, we compute the count statistics $b_{r,i'k}(\hat{\be}^{(s)})$ according to \eqref{eq: count} for all $i'=1,\cdots, n, k=1,\cdots, K$. For consistency, let $b_{r,i'k}^{(s)}= b_{r,i'k}(\hat{\be}^{(s)})$ and $\bb_{r,i'}^{(s)}=\{b_{r,i'k}^{(s)}\}_{k=1}^{K}.$
\item {\bf Step 2} (Update parameter estimates). In the $(t+1)$th step of EM iteration, we update $(\hat{\bpi}^{(s,t)}, \hat{\bLambda}^{(s,t)})$ by the following two steps: (1) E-step: estimate the probabilities of node labels by
\beq\label{eq: probability}
\hat{\tau}_{r,i'l}^{(s,t+1)}= P(z_{r,i'}=l| \bb^{(s)}_{r,i'})= \frac{\hat{\pi}^{(s,t)}_{l}\prod_{m=1}^{K} \exp{ (b^{(s)}_{r,i'm}\log{\hat{\lambda}^{(s,t)}_{lm}}- \hat{\lambda}^{(s,t)}_{lm})}}{\sum_{k=1}^{K}\hat{\pi}^{(s,t)}_{k}\prod_{m=1}^{K}\exp{(b^{(s)}_{r,i'm}\log{\hat{\lambda}^{(s,t)}_{km}}- \hat{\lambda}^{(s,t)}_{km})}},
\eeq
where $\hat{\tau}^{(s,t+1)}_{r,i'l}$ represents the estimated conditional probability that the node $i'$ belongs to the $l$th cluster, for all $1\le i' \le n, 1\le l\le K$; (2) M-step: given the label probabilities, we update the parameter estimates as follows,
\beq\label{eq: parameter}
\hat{\pi}^{(s,t+1)}_{l}= \frac{1}{n}\sum_{i'=1}^{n} \hat{\tau}^{(s,t+1)}_{r,i'l}, ~~\mbox{and}~~\ \hat{\lambda}^{(s,t+1)}_{lk}= \frac{ \sum_{i'=1}^{n}\hat{\tau}^{(s,t+1)}_{r,i'l}b^{(s)}_{r,i'k}}{ \sum_{i'=1}^{n} \hat{\tau}^{(s,t+1)}_{r,i'l}}.
\eeq
Repeat steps (1) and (2) until the EM algorithm converges.
\item {\bf Step 3} (Update label estimator). The estimates $(\hat{\bpi}^{(s+1)}, \hat{\bLambda}^{(s+1)})$ and $\{ \hat{\tau}^{(s+1)}_{r,i'l}\}$ are obtained from the last step of EM algorithm. We then update the in-worker node labels by $\hat{e}^{(s+1)}_{r,i'}= \argmax_{l} \hat{\tau}^{(s+1)}_{r,i'l}$, for $i'=1,\cdots, n$.
\end{itemize}

For illustration, we show the above procedures in Figure \ref{fig: worker-estimation}. Through the above iterations, each worker obtains an updated label estimator $\hat{\be}^{(s+1)}_{r}$ for in-worker nodes. This local estimation procedure has the following advantages. First, each worker identifies the community labels of the in-worker nodes using all relevant observed connectivities. This mechanism ensures the desirable efficiency of the distributed estimator. Second, the label estimates from the different workers are naturally aligned. For each worker, the update of the local labels $\hat{\be}_r$ matches the initial label estimator $\hat{\be}$. We then explain this phenomenon in detail with the following toy example.

\begin{figure}[!htb]
\centering
\includegraphics[width=0.70\linewidth]{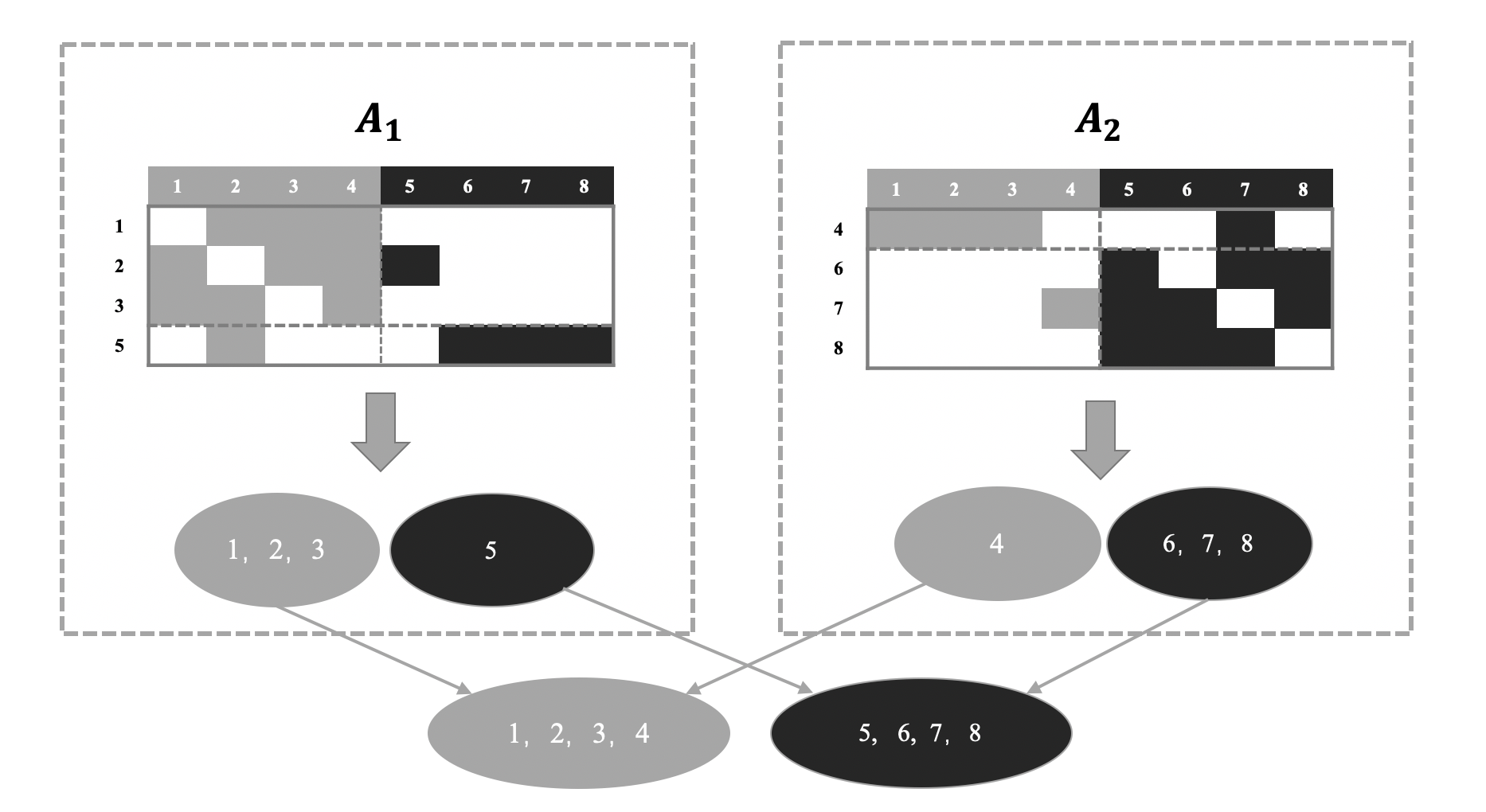}
 \caption{Community detection based on two sub-adjacency matrices, where white denotes $a_{ij}=$0 and black or gray denotes $a_{ij}=$1. The initial label assignments are represented by gray and black in the column indices of the sub-adjacency matrix. Specifically, we assume that gray and black denote clusters 1 and 2, respectively. Moreover, the label estimation results of in-worker nodes are shown via gray and black circles. Combining the circles with the same color, we could obtain the global estimator.}\label{fig: aligned}
\end{figure}

Specifically, in Figure \ref{fig: aligned}, the sub-adjacency matrices are obtained by the block-wise splitting of the entire network, as shown in the left panel of Figure \ref{fig: community}. For the first worker, the in-worker node set is $\{1,2,3,5\}$ and the initial label estimator assigns nodes \{1,2,3,4\} to cluster 1 and \{5,6,7,8\} to cluster 2. Based on $\bA_1$, we observe that nodes \{1,2,3\} and 5 have denser connections with clusters 1 and 2, respectively. Thus, according to the initial label assignments, the first local estimator assigns nodes \{1,2,3\} to cluster 1 and 5 to cluster 2. Moreover, the second local estimator assigns node 4 to 1, and nodes \{6,7,8\} to cluster 2. As a result, combining the two local label estimators allows us to obtain a global label estimator easily, resulting in a distributed framework that is both communication and computationally efficient. We then describe the distributed framework of community detection for the entire network in the following section.

\subsection{Distributed Pseudo-Likelihood Algorithm}

The distributed community detection framework is formed by a two-step communication between the master and the workers. Furthermore, to refine the estimation efficiency, we develop a multi-round iteration algorithm to fit the parameters in SBM. Specifically, in the $(s+1)$th iteration, we update $(\hat{\bpi}^{(s)}, \hat{\bLambda}^{(s)})$ and $\hat{\be}^{(s)}$ using a two-step communication between the master and the workers.

\begin{figure}[!h]
\centering
 \includegraphics[width=1.0\linewidth]{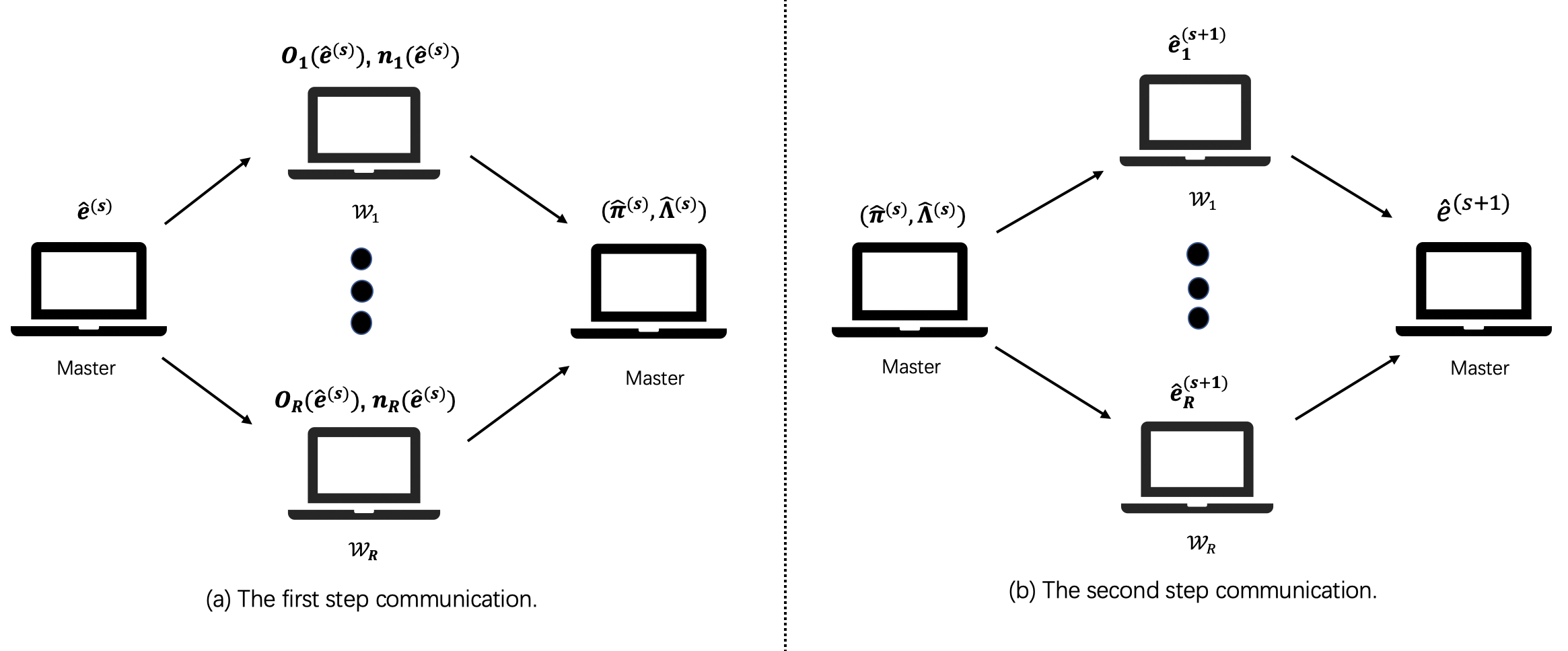}
 \caption{An example of two-step communication between $R$ workers and one master. Specifically, in the $(s+1)$th iteration, the distributed algorithm updates label vector $\hat{\be}^{(s)}$ using this two-step communication procedure. In the first-step communication, the master obtains the initial estimates $(\hat{\bpi}^{(s)}, \hat{\bLambda}^{(s)})$ for EM iteration. In the second-step communication, the master transmits the estimates to each worker to update the label vector of each in-worker node (i.e., $\hat{\be}_r^{(s+1)}, r=1,\cdots, R$). Finally, the master updates the label vector $\hat{\be}^{(s+1)}$ by aggregating the worker estimates.}\label{fig: two_step}
\end{figure}

In the first step, we obtain a global estimator of parameters $(\hat{\bpi}^{(s)}, \hat{\bLambda}^{(s)})$ based on the current label vector $\hat{\be}^{(s)}$. In each worker $\mW_r$, we compute some count statistics from sub-adjacency $\bA_{r}$, including the number of connectivities between the $l$th row cluster and the $k$th column cluster $O_{r,lk}(\hat{\be}^{(s)})= \sum_{i'=1}^{n}\sum_{j=1}^{N}a_{r,i'j}\bI(\hat{e}^{(s)}_{r,i'}=l, \hat{e}^{(s)}_{j}=k)$, and the number of nodes in the $l$th row cluster $n_{r,l}(\hat{\be}^{(s)})= \sum_{i'=1}^{n}\bI(\hat{e}^{(s)}_{r,i'}=l)$. We then transmit these summary statistics to the master to compute the estimates as
\beq\label{eq: initial_parameter}
\hat{\pi}^{(s)}_{l}=\frac{\sum_{r=1}^{R} n_{r,l}(\hat{\be}^{(s)})}{N}, ~~\mbox{and}~~\ \hat{\lambda}^{(s)}_{lk}= \frac{\sum_{r=1}^{R} O_{r,lk}(\hat{\be}^{(s)}) }{\sum_{r=1}^{R}n_{r,l}(\hat{\be}^{(s)})}.
\eeq
The parameter estimates $(\hat{\bpi}^{(s)}, \hat{\bLambda}^{(s)})$ and $\hat{\be}^{(s)}$ for each worker as the initial estimators in the EM iteration.

In the second step, we update the current label vector $\hat{\be}^{(s)}$. Specifically, we first update the labels of the in-worker node $\hat{\be}^{(s+1)}_{r}$ in parallel and transmit them to the master. We then combine these estimates to update the global estimator. Based on the block-wise splitting method, the network nodes $\mN$ are randomly divided into $R$ equal blocks; thus, the indices of the nodes are rearranged. To match the global index, for each node $i \in \mN$, assume $r_i$ as its block label, and define its index in the $r_i$th worker as $w_i$, where $1\le r_i \le R$ and $1\le w_i\le n$. Thus, we update the global estimator by considering $\hat{\be}^{(s+1)}_{i}=\hat{e}^{(s+1)}_{r_i,w_i}$ for $1\le i \le N.$

We repeat the above two steps until the pseudo log-likelihood converges, and obtain the estimators from the multi-round distributed computing. We refer to this multi-round two-step communication procedure as distributed pseudo-likelihood algorithm (DPL). Furthermore, we show two-step communication in Figure \ref{fig: two_step} and present the procedure of the DPL method in Algorithm \ref{alg: DPL} in detail.

\begin{algorithm}[!h]
\caption{Distributed Pseudo-Likelihood Algorithm (DPL)}
\begin{algorithmic}
\STATE \textbf{Step 1}: Initialize $\hat{\be}^{(0)}$ using spectral clustering with permutations (SCP, \citealt{amini2013pseudo}) to the first sub-adjacency matrix $\bA_{1}$ and distribute to workers;
\STATE \textbf{Repeat}
\begin{itemize}
\STATE \textbf{Step 2}: Each worker calculates $\bO_{r}(\hat{\be}^{(s)})$, $\bn_{r}(\hat{\be}^{(s)})$, and transmits to master;
\STATE \textbf{Step 3}: Master calculates $(\hat{\bpi}^{(s)},\hat{\bLambda}^{(s)})$ according to \eqref{eq: initial_parameter} and broadcasts to workers;
\STATE \textbf{Step 4}: Each worker computes the count statistics $\{\bb^{(s)}_{r,i'}\}_{i'=1}^{n}$ and then initializes $\hat{\bpi}^{(s,0)}=\hat{\bpi}^{(s)}$ and $ \hat{\bLambda}^{(s,0)}=\hat{\bLambda}^{(s)}$;
\STATE \textbf{Repeat}
\begin{itemize}
 \STATE \textbf{E-step}: each worker computes $\hat{\tau}^{(s,t+1)}_{r,i'l}$ using \eqref{eq: probability} for $1\le i' \le n$ and $1\le l \le K$;
 \STATE \textbf{M-step}: each worker calculates $\hat{\pi}^{(s,t+1)}_{r,l}$ and $\hat{\lambda}^{(s,t+1)}_{r,lk}$ using \eqref{eq: parameter} for $1\le k,l \le K$;
\end{itemize}
\STATE \textbf{Until} the EM algorithm converges;
\STATE \textbf{Step 5}: Each worker updates $\hat{e}^{(s+1)}_{r,i'}= \argmax_{l}\hat{\tau}^{(s+1)}_{r,i'l}$, for all $1\le i' \le n$, and transmits $\hat{\be}^{(s+1)}_{r}$ to master;
 \STATE \textbf{Step 6}: Master updates the global estimator $\hat{e}^{(s+1)}_{i}=\hat{e}^{(s+1)}_{r_i,w_i}$ for $1\le i \le N$, and broadcasts to workers;
\end{itemize}
\STATE \textbf{Until} the pseudo log-likelihood converges.
\end{algorithmic}\label{alg: DPL}
\end{algorithm}

\begin{remark}[Determine the number of communities]\label{remark: determine} For real-world datasets with an unknown number of communities, the corrected Bayesian information criterion proposed by \cite{hu2020corrected} is adopted to determine the number of clusters $K$ for the proposed method in a distributed way. Let $\mK$ denote the candidate set for the number of communities, then we evaluate each candidate $K' \in \mK$ by three steps: (1) use the DPL method to estimate the corresponding label vector $\hat{\be} \in [K']^{N}$ and transmit $\hat{\be}$ to each worker; (2) each worker calculates the log-likelihood of $\bA_r$ by $\ell(K', \hat{\be},\bA_r)= \sum_{l=1}^{K'}\sum_{k=1}^{K'} \big[O_{r,lk}(\hat{\be})\log{\{\hat{\theta}_{r,lk}/(1- \hat{\theta}_{r,lk})\}}- n_{r,lk}(\hat{\be}) \log{(1- \hat{\theta}_{r,lk})}\big],$ where $\hat{\theta}_{r,lk}=\hat{\lambda}_{r,lk}(\hat{\be})/\sum_{i=1}^{N}\bI(\hat{\be}_{i}=k)$; (3) the master calculates the corrected Bayesian information criterion by $L(K',\hat{\be},\bA)=\sum_{r=1}^{R} \ell(K', \hat{\be},\bA_r) -\{ N \log{K'}+K'(K'+1)/2\log{ N}\}.$ As a result, the optimal solution is $\hat{K}=\argmax_{K'\in \mK} L(K', \hat{\be},\bA).$
\end{remark}

\subsection{Theoretical Discussions of the DPL Algorithm}

In a distributed system, a large number of workers corresponds to a small sample size for each worker, which yields higher computational efficiency. However, because each worker only has access to a local sample, the estimation accuracy is undesirable in this case. Therefore, we focus on establishing a theoretical lower bound for the worker sample size $n$.

We first introduce some necessary notations and assumptions. Define $\pi_{\rm min}=\min_{k} \pi_k$, where $N\pi_{\rm min}$ denotes the minimum community size. Then, the following assumptions are needed.
\begin{enumerate}
\item (Balanced level) Assume that there exists a positive constant $c>0$ such that $\pi_{\rm min} \ge c$.\label{ass: balance}
\item (Homogeneity of subnetworks) Assume that the sub-adjacency matrices $(\bA_r)_{r=1}^R$ are independent identically distributed variables.\label{ass: homogeneity}
\item (Network density) Assume the connectivity matrix $\bTheta=\rho \bTheta^{*}$, where $\bTheta^{*}\in (0,1)^{K\times K}$ is a constant matrix and $\rho \to 0$ at a rate of $N\rho=\Omega(\log{N})$.\label{ass: density}
\end{enumerate}
\noindent Assumption \ref{ass: balance} implies that the size of each community grows to infinity at the same rate as the total network size $N$. This is a mild condition for large-scale networks and it is also considered by \cite{amini2013pseudo} and \cite{wang2021fast}. Under SBM, Assumption \ref{ass: homogeneity} is easy satisfied if the node set $\mN$ is randomly divided into $R$ relatively uniform subsets. This assumption is adopted based on the conditions in \cite{jordan2019communication} and \cite{fan2021communication}. Assumption \ref{ass: density} is used to constraint the network density, which allows a sparse network, and is also adopted by \cite{chaudhuri2012spectral} and \cite{lei2015consistency}.

We provide two necessary conditions for the subnetwork for each worker to efficiently identify node labels. First, for each worker, in-worker nodes are required to cover all communities with a high probability.
Specifically, under the SBM model with $K$ blocks, we define a set of node sets that cover $K$ blocks completely as $\mS_{K}=\{S : \forall \ k\in \{1,\cdots, K\}, \ \exists \ i \in S \ \text{s.t.,} \ \bz_{i}=k\}$. Second, we require that the average degree of the subnetwork for each worker grows with the entire network size $N$. Specifically, let $d_{r,j}= \sum_{i'=1}^{n} a_{r,i'j}$ denote the degree of node $j$ in the $r$th subnetwork. Moreover, let $d_r=\sum_{j=1}^{N}d_{r,j}/N$ represent the average degree of the subnetwork.
Then, we constrain the expected average degree $E(d_r)=\Omega(\log{N})$. In other words, there exists a constant $c$ and a positive $N_0$ such that $E(d_r)>c\log{N}$ for all $N\ge N_0$. Based on these two conditions, we provide the lower bound of the worker sample size in the following theorem.

\begin{theorem}[\textbf{Worker sample size}]\label{the: sampleSize} Consider a network generated from the SBM with $K$ blocks. Under Assumptions \ref{ass: balance}--\ref{ass: density}, if the worker sample size $n= \Omega(\log{N}/\rho)$, then for each subsample $\mN_r$, we have $\mN_r \in \mS_{K}$ and $E(d_r)=\Omega(\log{N})$ with probability at least $1-1/N$.
\end{theorem}
\noindent Technique proof of this theorem is provided in Appendix A.1. According to Theorem \ref{the: sampleSize}, if the network density $\rho=(\log{N})^{-1}$, then subsample size on each worker can be of the order $O\{(\log{N})^2\}$. It is remarkable that this restriction is milder than that in one-shot distributed computing (i.e., $n=O(\sqrt{N})$, \citealt{zhang2013communication}).

Based on the conclusion of Theorem \ref{the: sampleSize}, we discuss the computational complexity of each iteration of the DPL algorithm in the following proposition.
\begin{proposition}[\textbf{Computational complexity}]\label{pro: computational} Assume that the entire network is evenly divided by the block-wise method and each subnetwork has $n$ in-worker nodes. Hence, the computational complexity per iteration of the DPL algorithm is $O(Nn\rho_N)$, where $\rho_N$ is the network density.
\end{proposition}
\noindent The proof of this proposition is given in Appendix A.2. According to Proposition \ref{pro: computational}, since the worker sample size $n \ll N$, the computational complexity of DPL is much lower than that of existing distributed/parallel community detection methods \citep{chen2010parallel, su2021parallel, mukherjee2021two, wu2023distributed}. Specifically, under the conditions in Theorem \ref{the: sampleSize}, the computational complexity of the proposed method could be $O(N\log{N}).$

We then provide the upper bound of the communication cost in the following proposition. Notably, the DPL algorithm consists of multiple rounds of communication between the master and the workers, and we next discuss the communication cost of DPL in each iteration, namely the two-step communication.
\begin{proposition}[\textbf{Communication cost}]\label{pro: communication} Under the same assumptions of Proposition \ref{pro: computational}, the communication cost per iteration of DPL is $O(NR)$ bits, where $R=N/n$.
\end{proposition}
\noindent The proof of this proposition is provided in Appendix A.3. According to Proposition \ref{pro: communication}, the communication cost increases with the number of workers (i.e., $R$). The most expensive communication is the first-step communication, wherein the master broadcasts the initial global estimator to each worker. In the second communication step, each worker sends its local label estimator to the master, which only requires $O(N/R)$ bits per worker. The method proposed by \cite{chen2010parallel} also requires $O(NR)$ bits for communication.

\subsection{Extension to the Degree-Corrected SBMs}

Recall that the DCSBM is a generalization model of the SBM that allows degree heterogeneity within communities. Therefore, to better fit the real-world network, we investigated the development of a distributed algorithm to identify the community structure of networks generated from DCSBMs. According to \cite{jin2015fast}, the degree parameters are not useful for identifying community structures. Even worse, when some nodes have a large degree, they can cause computer memory errors when computing pseudo-likelihood. Therefore, we use the conditional likelihood method to carefully eliminate their influence.

Specifically, for any node $i'$ in the $r$th worker, denote its node degree as $d_{r,i'}=\sum_{j=1}^{N}a_{r,i'j}$, and thus we have $d_{r,i'}=\sum_{k=1}^{K}b_{r,i'k}$. In this way, conditional on node degree $d_{r,i'}$ and node label $z_{r,i'}=l$, the count statistics $\{b_{r,i'k}\}_{k=1}^{K}$ are multinomial variables with parameters $(\psi_{l1},\cdots, \psi_{lK})$, where $\psi_{lk}=\lambda_{lk}/\lambda_l.$ Therefore, the conditional pseudo log-likelihood function of $\{\bb_{r,i'}\}_{i'=1}^{n}$ is given by
\beq\label{eq: cpl}
\ell_{\rm DCSBM}(\pi, \bPsi; \{\bb_{r,i'}\}_{i'=1}^{n})= \sum_{i'=1}^n \log\Big\{ \sum_{l=1}^K \pi_{l} \prod_{k=1}^{K}\psi_{lk}^{b_{r,i'k}}\Big\},
\eeq
where $\bPsi=(\psi_{lk}) \in (0,1)^{K\times K}$. Similar to the unconditional case, we use an iterative algorithm to estimate the parameters based on \eqref{eq: cpl}.

\begin{algorithm}[!h]
\caption{Distributed Conditioned Pseudo-Likelihood Algorithm (DCPL)}
\begin{algorithmic}
\STATE \textbf{Step 1}: Initialize $\hat{\be}^{(0)}$ using the spherical spectral clustering algorithm (SSC, \citealt{lei2015consistency}) to the first sub-adjacency matrix $\bA_{1}$;
\STATE \textbf{Repeat}
\begin{itemize}
\STATE \textbf{Step 2}: Each worker calculates $\bO_{r}(\hat{\be}^{(s)})$ and $\bn_{r}(\hat{\be}^{(s)})$ and transmits to master;
\STATE \textbf{Step 3}: Master calculates $(\hat{\bpi}^{(s)},\hat{\bLambda}^{(s)})$ according to \eqref{eq: initial_parameter} and then computes $\hat{\bPsi}^{(s)}$ by $\hat{\psi}_{lk}^{(s)}= \hat{\lambda}^{(s)}_{lk}/\sum_{m=1}^K\hat{\lambda}^{(s)}_{lm},$ and then broadcasts $(\hat{\bpi}^{(s)}, \hat{\bPsi}^{(s)})$ to workers;
\STATE \textbf{Step 4}: Each worker computes the count statistics $\{\bb^{(s)}_{r,i'}\}_{i'=1}^{n}$ corresponding to $\hat{\be}^{(s)}$ and then initializes $\hat{\bpi}^{(s,0)}=\hat{\bpi}^{(s)}$ and $\hat{\bPsi}^{(s,0)}=\hat{\bPsi}^{(s)}$;
\STATE \textbf{Repeat}
\begin{itemize}
\STATE \textbf{E-step}: each worker computes $\hat{\tau}^{(s,t+1)}_{r,i'l}$ using \eqref{eq: cond-probability} for $1\le i' \le n$ and $1\le l \le K$;
\STATE \textbf{M-step}: each worker calculates $\hat{\pi}^{(s,t+1)}_{r,l}$ and $\hat{\psi}^{(s, t+1)}_{r,lk}$ using \eqref{eq: cond-parameter} for $1\le k,l \le K$;
\end{itemize}
\STATE \textbf{until} the EM algorithm converges;
\STATE \textbf{Step 5}: Each worker updates $\hat{e}^{(s+1)}_{r,i'}= \argmax_{l}\hat{\tau}^{(s+1)}_{r,i'l}$, for all $1\le i' \le n$ and then transmits $\hat{\be}^{(s+1)}_{r}$ to the master;
 \STATE \textbf{Step 6}: Master updates the global estimator $\hat{e}^{(s+1)}_{i}= \hat{e}^{(s+1)}_{r_i,w_i}$ for $1\le i \le N$, and broadcasts to workers;
\end{itemize}
\STATE \textbf{Until} the conditional pseudo log-likelihood converged.
\end{algorithmic}\label{alg: DCPL}
\end{algorithm}

The update procedure for each worker is the same as that for the unconditional pseudo log-likelihood, with Step 2 replaced by
\begin{itemize}
\item {\bf Step $2'$} (Update parameter estimates). In the $(t+1)$th step of EM iteration, we update $(\hat{\bpi}^{(s,t)}, \hat{\bPsi}^{(s,t)})$ by the following two steps:
\begin{itemize}
\beq\label{eq: cond-probability}
\hat{\tau}_{r,i'l}^{(s,t+1)}= P(z_{r,i'}=l| \bb^{(s)}_{r,i'})= \frac{\hat{\pi}^{(s,t)}_{l} \prod_{m=1}^K \{\hat{\psi}_{lm}^{(s,t)}\}^{b^{(s)}_{r,i'm}} }{ \sum_{k=1}^K \hat{\pi}^{(s,t)}_k \prod_{m=1}^K \{\hat{\psi}^{(s,t)}_{km}\}^{b^{(s)}_{r,i'm}}}.
\eeq
\item[(2)] M-step: based on the label probabilities, we have
\beq\label{eq: cond-parameter}
\hat{\pi}_{l}^{(s,t+1)}= \frac{1}{n}\sum_{i'=1}^n \hat{\tau}^{(s,t+1)}_{r,i'l}, ~~\mbox{and}~~\ \ \hat{\psi}^{(s,t+1)}_{lk}= \frac{\sum_{i'=1}^{n} \hat{\tau}_{r,i'l}^{(s,t+1)} b^{(s)}_{r,i'k}}{\sum_{i'=1}^{n} \hat{\tau}^{(s,t+1)}_{r,i'l}d_{r,i'}}.
\eeq
\end{itemize}
\end{itemize}

Based on the calculation of each worker, we construct a distributed conditional pseudo-likelihood (DCPL) algorithm for DCSBM in Algorithm \ref{alg: DCPL}. Similar to Algorithm \ref{alg: DPL}, this distributed framework is formed by multiple rounds of two-step communication, with the difference being in Steps 1, 3, and 4. In Step 1, under the DCSBM framework, we apply the spherical spectral clustering (SSC) proposed by \cite{lei2015consistency} to obtain the initial labels. In Step 3, because we use the conditional distribution of the count statistics, the master normalizes each row of matrix $\hat{\bLambda}^{(s)}$ to obtain $\hat{\bPsi}^{(s)}$. Furthermore, Step 4 is accomplished by each worker using the EM algorithm for conditional pseudo log-likelihood.

\section{Numerical Studies}

\subsection{Simulation Models and Performance Measurements}

We evaluate the performance of the DPL method using the following three examples. First, we examine the consistency of the DPL method in identifying the community labels of the network nodes. Second, we investigate the effect of community signal strength on the performance of the proposed method. Finally, we evaluate the performance of the DPL method in heterogeneous networks.

\begin{example}[Consistency of clustering results]\label{exam: consistency} Assume the connectivity matrix $\bTheta =\rho\{(1-\beta) \one_K \one_K^{\top} + \beta {\bm I}_{K}\}$, where $\one_K$ is filled with elements 1,  $0\le \rho, \beta \le 1$, and ${\bm I}_{K}\in\mR^{K\times K}$ is an identity matrix. Specifically, let $K=3$ and $\bpi= (0.2,0.3, 0.5)$ and assign each worker an equal sample size $n$. Thereafter, two different cases are considered: (1) set $\rho=5\times 10^{-3}, \beta =0.8$ and fix the total network size $N=10,000$, and let worker sample size $n$ vary from 100 to 1,000; (2) set $\rho=3\times 10^{-3}, \beta=0.8$ and fix the worker sample size $n$ at 200, and let the total network size $N$ increase from 2,000 to 30,000.
\end{example}

\begin{example}[Effect of signal strength]\label{exam: signal} The connectivity matrix $\bTheta$ is assumed to be the same as Example \ref{exam: consistency}. Moreover, we fix $N=10,000$, $n=500$, $K=3$, and $\bpi= (0.2,0.3, 0.5)$. Under the SBM, the signal strength of the community structure depends on the connectivity density $\rho$ and the connectivity divergence $\beta$. Here, we consider two cases: (1) with a fixed $\beta=0.8$, let $\rho$ increase from 0.001 to 0.01; (2) with a fixed $\rho=0.01$, $\beta$ varies from 0.1 to 0.9.
\end{example}

\begin{example}[Degree heterogeneous network]\label{exam: degree} The degree parameters $\{\alpha_i\}_{i=1}^N$ are generated according to \cite{zhao2012consistency}. Specifically, define $P(\alpha_i= mx)= P(\alpha_i =x)= 1/2, \ \text{with} \ x= 2/(m+1),$ which ensures $E(\alpha_i)=1$. We set $m$ to increase from 2 to 10, where a larger $m$ corresponds to a higher degree of heterogeneity. Moreover, we fix $N=10,000$, $n=500$, $K=3$, and let $\bTheta= 3\times 10^{-3} \{ \one_{K}\one_K^{\top} + \diag(2,3,4)\}$. We then consider the performance evaluated at relatively balanced community sizes with $\bpi=(0.3,0.3,0.4)$ and imbalanced community sizes with $\bpi=(0.1,0.2,0.7)$, respectively.
\end{example}

{\sloppy To measure the performance of the proposed method, we consider the widely used criterion of normalized mutual information (NMI). For any community label estimator $\hat{\be}$, we define a $K\times K$ confusion matrix $\bM$ with $M_{kl}= 1/N\sum_{i=1}^{N}\bI(\hat{e}_i=k, z_i=l)$ for $1\le k, l\leq K$. Additionally, we denote the row and column marginal sums as $\bM_{l\cdot}$ and $\bM_{\cdot k}$, respectively. Then, the NMI is defined by \cite{yao2003information} as ${\rm NMI}(\bz,\hat{\be})= -\sum_{l,k}\bM_{lk}\log{\frac{\bM_{lk}}{\bM_{l\cdot}\bM_{\cdot k}}(\sum_{l,k}\bM_{lk} \log{\bM_{lk}} )^{-1}}$. The NMI value is supposed to be between 0 and 1, where a larger NMI indicates better clustering performance.}

A comparison of the methods is presented in Table \ref{tab: methods}. The pseudo-likelihood method (PL) is proposed based on SBM assumptions, employing the pseudo-likelihood method as its objective function and using the EM algorithm to derive the label estimator. The conditional pseudo-likelihood (CPL) is an extension of the PL method, developed under DCSBM assumptions. Both methods are performed on a single machine, using the entire network data. The parallel spectral clustering (PSC) method implements the spectral clustering algorithm in a divide-and-conquer fashion. Furthermore, \cite{mukherjee2021two} proposed two methodologies: the piecewise averaged community estimation (PACE) and the global alignment of local estimates (GALE) methods. In these approaches, each worker performs community detection on the corresponding subgraph. In PACE, the master derives the global label estimator by averaging local clustering matrices, while in the GALE method, the master sequentially matches the local estimates based on their confusion matrix to obtain the global label estimator.

It is worth noting that the PACE and GALE methods are conditioned on the local subsample size, and in our simulation experiments, we set the subsample size in PACE and GALE to be five times larger than that of the proposed DPL and DCPL methods. Each random experiment is repeated 100 times to ensure reliable simulation results. To ensure a fair comparison, all algorithms are implemented in Python 3.10. All simulations are conducted on a Linux server equipped with an Intel Xeon E5-2650 v4 CPU, boasting 24 cores and 64GB of RAM.

\begin{table}[ht]
\centering
\caption{Comparison of community detection algorithms.}\label{tab: methods}
\begin{tabular}{cccc}
  \hline
  Model & Method & \makecell[c]{Distributed \\ computing} & \makecell[c]{Network \\ information}\\
  \hline
 SBM  & PL \citep{amini2013pseudo} & No & Entire network \\
  & PSC \citep{chen2010parallel} & Yes & Entire network \\
  & PACE \citep{mukherjee2021two} &Yes & Subnetwork\\
  & GALE \citep{mukherjee2021two} &Yes & Subnetwork\\
  & DPL & Yes & Entire network \\
   \hline
DCSBM & CPL \citep{amini2013pseudo} & No & Entire network \\
  & PSC \citep{chen2010parallel} & Yes & Entire network \\
  & PACE \citep{mukherjee2021two} &Yes & Subnetwork\\
  & GALE \citep{mukherjee2021two} &Yes & Subnetwork\\
 & DCPL & Yes & Entire network \\
   \hline
\end{tabular}
\end{table}

\subsection{Simulation Results}

\begin{figure}[!htb]
   \centering
   \includegraphics[width=1.0\textwidth]{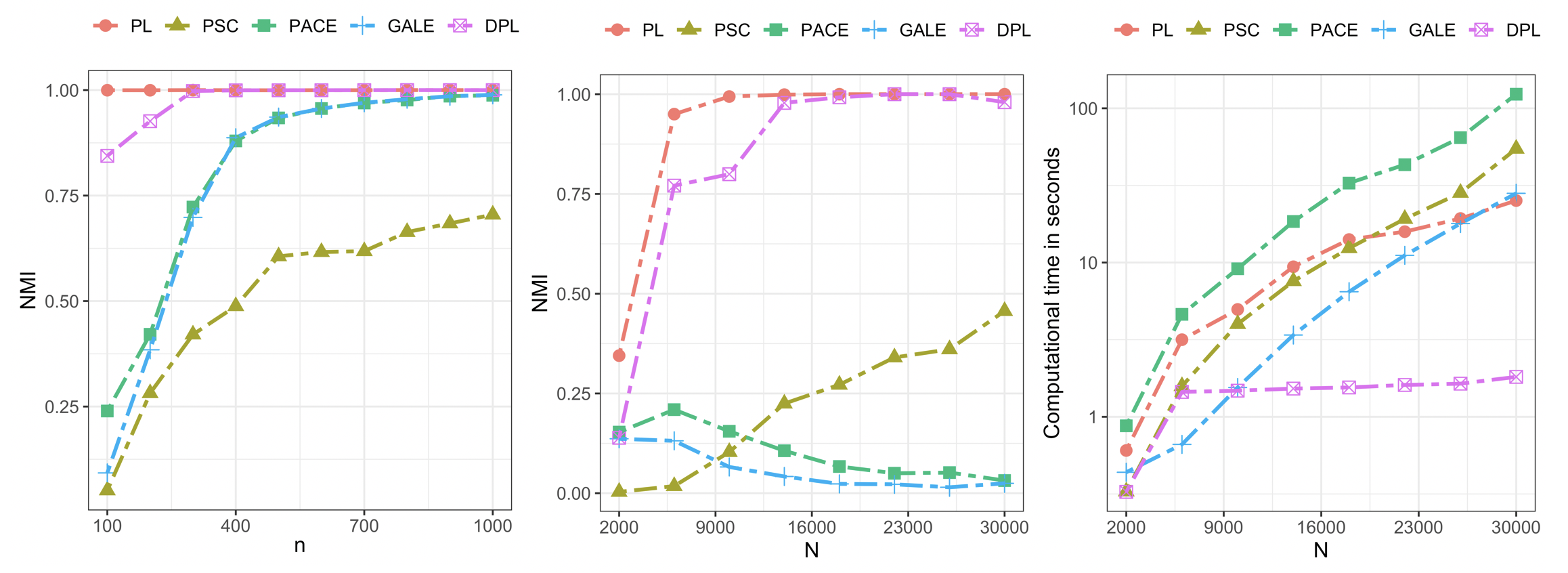}
   \caption{\small Simulation results for Example \ref{exam: consistency}. In the left panel, the worker sample size increases from 100 to 1,000 while the total network size is fixed at $N=10,000$. In the middle and right panels, the total network size varies from 2,000 to 30,000 while each worker sample size is fixed at $n=200$. The computational time of these algorithms is compared in the right panel, where the y-axis is presented in a logarithmic scale.}\label{fig: consistency}

\end{figure}

\begin{figure}[!htb]
   \centering
    \includegraphics[width=0.68\textwidth]{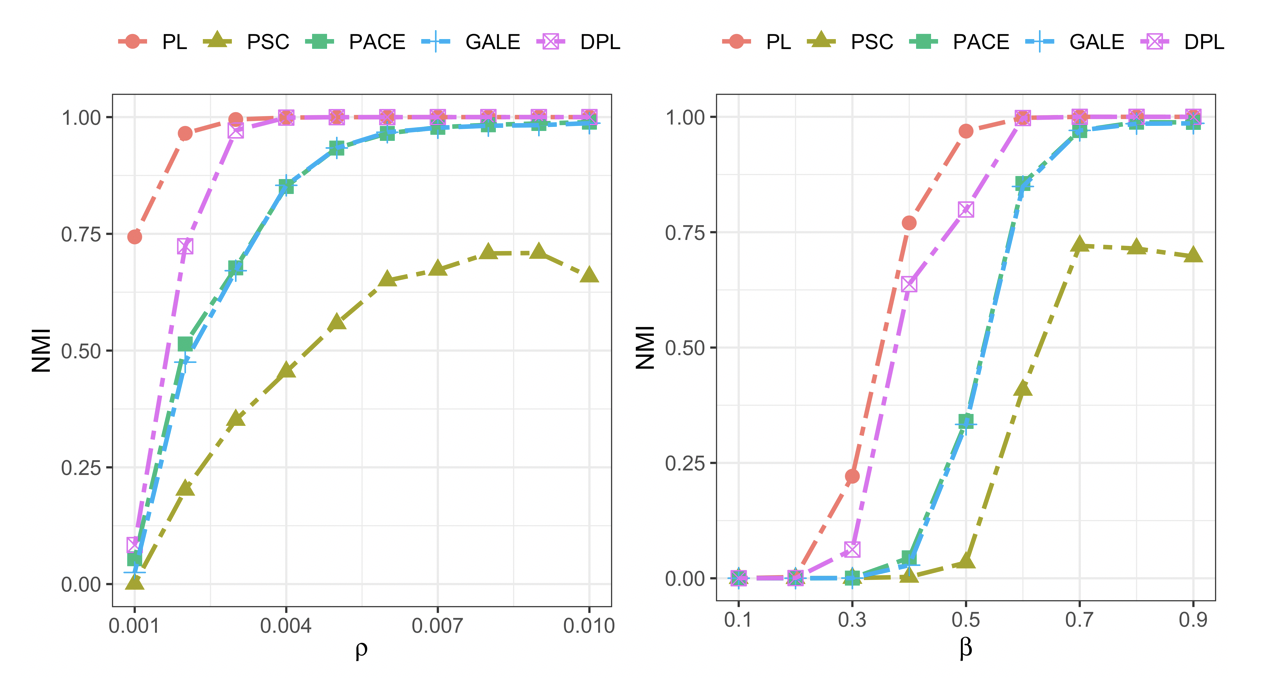}
        \caption{\small Simulation results for Example \ref{exam: signal}. The effect of connectivity density $\rho$ and connectivity divergence $\beta$ on the performance of each community detection method.}\label{fig: signal}
\end{figure}

\begin{figure}[!htb]
   \centering
    \includegraphics[width=0.68\textwidth]{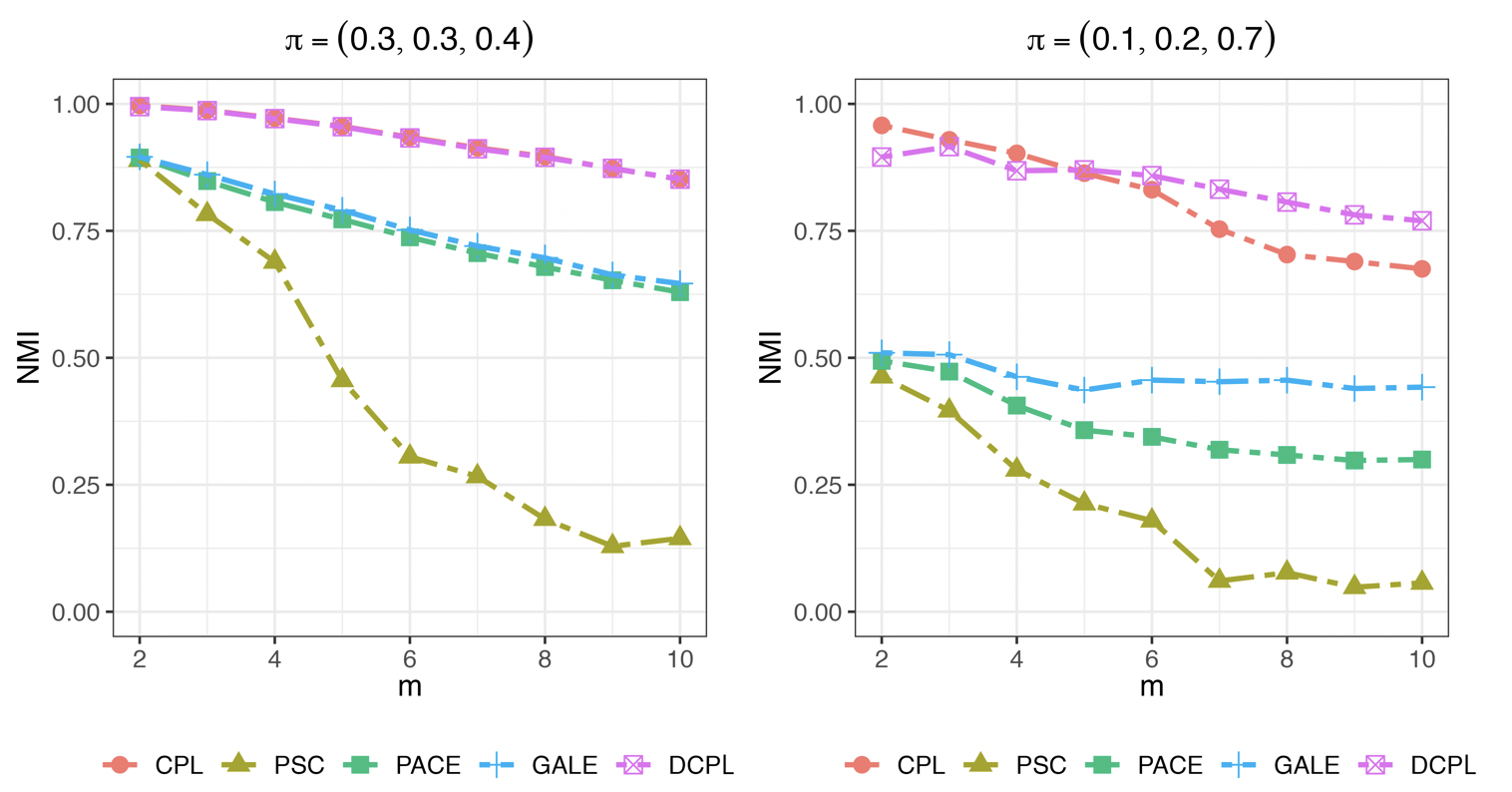}
     \caption{\small Simulation results for Example \ref{exam: degree}. The effect of degree heterogeneity on the performance of each community detection method, where the left and right panels show the community detection performance evaluated at relatively balanced (i.e., $\bpi=(0.3,0.3,0.4)$) and imbalanced (i.e., $\bpi=(0.1,0.2,0.7)$) community sizes, respectively.}\label{fig: degree}
\end{figure}

All the simulation results are shown in Figures \ref{fig: consistency}--\ref{fig: degree}. We draw the following conclusions from the three examples.

{\sc Example \ref{exam: consistency}.} The simulation results are presented in Figure \ref{fig: consistency}. First, as shown in the left panel of Figure \ref{fig: consistency}, as the worker sample size $n$ increases from 100 to 1,000, the NMI of the DPL method converges faster to 1.0 compared to the PSC, PACE, and GALE methods. This observation is consistent with the theoretical result stated in Theorem \ref{the: sampleSize}, which suggests that DPL requires a milder restriction on the worker sample size. Second, as shown in the middle panel of Figure \ref{fig: consistency}, with a fixed sample size $n=200$, the NMI of PL and DPL quickly reaches 1.0 as the total network size $N$ increases from 2,000 to 30,000, as expected. However, the clustering accuracy of the PACE and GALE methods decreases as more data becomes available owing to the fixed subsample size $n$, which becomes relatively smaller compared to the increasing $N$. Third, as shown in the right panel of Figure \ref{fig: consistency}, where the y-axis is presented on a logarithmic scale, as $N$ increases, the computational time of DPL only exhibits a slight increase, while that of other approaches grows dramatically. This arises from the DPL method performing the pseudo-likelihood approach within a distributed system, where each worker conducts lower-dimensional matrix operations to update their local estimator. Additionally, the master can easily combine these local estimators to derive the global estimator.

{\sc Example \ref{exam: signal}.} The simulation results are shown in Figure \ref{fig: signal}. First, for $\beta=0.8$, as the network density $\rho$ increases, the performance of PL and DPL exhibits significant improvement. Additionally, the accuracy of PSC, PACE, and GALE methods also increases with higher network density. The presence of multiple disconnected components in sparse networks poses challenges for community recovery. This suggests that PL and DPL methods are better suited for such scenarios due to their reliance on pseudo-likelihood of count statistics, offering greater robustness compared to spectral clustering-based approaches. Second, for $\rho=0.01$, as the connectivity divergence parameter $\beta$ increases from 0.1 to 0.9, the NMI of all compared algorithms increases. Notably, DPL achieves similar accuracy to PL and surpasses other algorithms. With the increase in $\beta$, the community structure of the entire network becomes more distinct.

{\sc Example \ref{exam: degree}.} The simulation results are shown in Figure \ref{fig: degree}. First, as the degree of heterogeneity $m$ increases from 2 to 10, the clustering accuracy of all methods decreases. However, the CPL and DCPL methods still exhibit better performance compared to the other methods, considering these methods effectively eliminate the influence of the degree parameter by utilizing the conditional pseudo-likelihood method. Additionally, these methods have the ability to fully leverage all available connection information. Second, comparing the left and right panels of Figure \ref{fig: degree}, all community detection methods perform better when community sizes are balanced. This observation can be attributed to the difficulty of community detection in small communities, where nodes tend to merge into larger communities during the community recovery process.

\subsection{Real Data Analysis}

Further, we assess the effectiveness of the proposed method through seven real data analysis examples obtained from the Stanford large network dataset collection \footnote{\url{http://snap.stanford.edu/data/}}. The selected datasets include ca-HepPh, ca-AstroPh, ca-CondMat, cit-HepPh, email-Enron, loc-Brightkite, and loc-Gowalla. The number of nodes in these networks ranges from 12,008 to 196,591, and detailed information on these real-world networks is provided in Table \ref{tab: networkProperties}.

{\sloppy In real applications, because the underlying true community labels are unknown, the community detection results are evaluated using the relative density \citep{chen2014community}. Specifically, given a label estimator $\hat{\be}$, we define the relative density as ${\rm RED}=C_{\rm between}(\bA, \hat{\be})/C_{\rm within}(\bA, \hat{\be}),$ where $C_{\rm between}(\bA, \hat{\be})= \sum_{i,j}a_{ij}\bI(\hat{\be}_i\neq \hat{\be}_j)/\sum_{i,j}\bI(\hat{\be}_i\neq \hat{\be}_j)$ is the between-community density and $C_{\rm within}(\bA, \hat{\be})= \sum_{i,j}a_{ij}\bI(\hat{\be}_i=\hat{\be}_j)/\sum_{i,j}\bI(\hat{\be}_i=\hat{\be}_j)$ is the within-community density. Thus, a small ${\rm RED}$ corresponds to a better network partition result. To ensure a fair comparison, all algorithms are implemented in Python 3.10 and executed on a Linux server with an Intel 6438M CPU, boasting 64 cores and 512GB of RAM. To illustrate computational complexity, we present the computational time in seconds for each algorithm.}

\begin{table}[!htb]
\centering
\caption{Properties of real-world networks.}\label{tab: networkProperties}
\begin{tabular}{cccc}
\hline
Network & Node number & Edge number & Network density \\
\hline
ca-HepPh & 12,008 & 118,521 & $1.01\times 10^{-3}$\\
ca-AstroPh & 18,772 & 198,110 & $1.01\times 10^{-5}$ \\
ca-CondMat & 23,133 & 93,497 & $3.49\times 10^{-4}$ \\
cit-HepPh & 34,546 & 421,578 & $7.07\times 10^{-4}$ \\
email-Enron & 36,692 & 183,831 & $2.73\times 10^{-4}$\\
loc-Brightkite & 58,228 & 214,078 & $1.26\times 10^{-4}$\\
loc-Gowalla & 196,591 & 950,327 & $4.92 \times 10^{-5}$\\
\hline
\end{tabular}
\end{table}

Considering the degree of heterogeneity, we fit these real-world networks using DCSBM. Therefore, in this experiment, we compare the proposed DCPL method with four aforementioned community detection algorithms: CPL \citep{amini2013pseudo}, PSC \citep{chen2010parallel}, PACE \citep{mukherjee2021two}, and GALE \citep{mukherjee2021two}. In addition, the comparison algorithms include the parallel multi-objective evolutionary algorithm (PMOEA) proposed by \citep{su2021parallel} and the distributed community detection method (DCD) developed by \cite{wu2023distributed}. The  PMOEA method utilizes evolutionary algorithms to address multiple objectives in community detection, and the DCD method is designed specifically for spectral clustering within a distributed system.

Before clustering real-world networks, we determine the number of clusters by the procedure in Remark \ref{remark: determine}, and the results are shown in Table \ref{tab: realData}. We use the corresponding number of clusters for all the compared and proposed methods for the sake of comparison. Additionally, to assess the impact of the subsample size, we vary the subsample size $n$ for the distributed algorithms, namely PSC, PACE, GALE, DCD, and DCPL. The subsample size is evenly distributed across multiple workers. Notably, the parallel algorithm, PMOEA, makes use of the entire network in each parallel computation. Detailed results are reported in Table \ref{tab: realData}. In the table, we use ``---'' to indicate instances where Python reports out-of-memory errors.

It is remarkable that for the loc-Gowalla network, all compared algorithms encounter out-of-memory errors when handling this large-scale network in our computing environment. In this way, we only report the computational time for the proposed DCPL method. The proposed DCPL method partitions this network into 12 clusters within 80.64 seconds for a subsample size of $n=3,000$ and 182.68 seconds for $n=5,000$. In comparison, in the study by \cite{su2021parallel}, the authors reported a computational time of 40,194 seconds for applying PMOEA to cluster the loc-Gowalla network using their computational resources. The community detection results for other datasets are shown in Table \ref{tab: realData}.

Based on the results presented in Table \ref{tab: realData}, several conclusions can be drawn. First, it can be seen that the proposed DCPL method achieves the best RED across all networks in comparison to distributed algorithms like PSC, PACE, GALE, and DCD. The superior performance of the DCPL method primarily stems from its ability to utilize connection information fully for label estimator updates. In contrast, PACE, GALE, and DCD identify community structures from subnetwork rather than the entire network during distributed calculations. Additionally, the PSC algorithm conducts parallel SVD to obtain low-dimensional node representations and imposes stricter conditions on subsample sizes in each worker to ensure clustering accuracy \citep{chen2010parallel}.

\begin{table}[!htb]
\caption{The report includes the relative density (RED) and computational time (CPT) presented in seconds for each compared algorithm. For each dataset, the best performance among the distributed algorithms under each subsample size setting is highlighted in bold text.}\label{tab: realData}
\resizebox{\columnwidth}{!}{
\begin{tabular}{cc rr| rrr| r}
\hline
                      & Network & \multicolumn{1}{c}{ca-HepPh} & \multicolumn{1}{c|}{ca-AstroPh} & \multicolumn{1}{c}{ca-CondMat} & \multicolumn{1}{c}{cit-HepPh} & \multicolumn{1}{c|}{email-Enron} & \multicolumn{1}{c}{loc-Brightkite}  \\
Method                & $K$       & \multicolumn{1}{r}{6}        & \multicolumn{1}{r|}{6}          & \multicolumn{1}{r}{6}          & \multicolumn{1}{r}{9}         & \multicolumn{1}{r|}{9}           & \multicolumn{1}{r}{12}                  \\
\multirow{2}{*}{CPL}  & RED     & 0.17                         & 0.15                           & 0.17                           & 0.03                          & 0.11                            & 0.10                                \\
                      & CPT     & 11.81                        & 19.67                          & 28.75                          & 56.18                         & 70.69                           & 115.40                         \\
          &   \multicolumn{1}{r}{}        & \multicolumn{1}{r}{}        & \multicolumn{1}{r|}{}          & \multicolumn{1}{r}{}          & \multicolumn{1}{r}{}         & \multicolumn{1}{r|}{}           & \multicolumn{1}{r}{}               \\
\multirow{2}{*}{PMOEA}   & RED     & 0.11                         & 0.51                           & 0.14                           & 0.04                          & 0.15                            & ---                                \\
                      & CPT     & 2332.42                      & 6170.76                        & 7270.65                        & 17776.56                      & 23002.37                        & ---                              \\
                      \hline
 &         & \multicolumn{2}{c|}{$n=500$} & \multicolumn{3}{c|}{$n=1,000$}      & \multicolumn{1}{c}{$n=3,000$}                                    \\
  \cline{3-4}  \cline{5-7}       \cline{8-8}
\multirow{2}{*}{PSC}  & RED     & 0.81                         & 0.91                           & 0.89                           & 0.58                          & 1.05                            & 0.39                               \\
                      & CPT     & 4.96                         & 9.70                           & 14.81                          & 32.26                         & 35.67                           & 115.44                             \\
  &   \multicolumn{1}{r}{}        & \multicolumn{1}{r}{}        & \multicolumn{1}{r|}{}          & \multicolumn{1}{r}{}          & \multicolumn{1}{r}{}         & \multicolumn{1}{r|}{}           & \multicolumn{1}{r}{}               \\
\multirow{2}{*}{PACE} & RED     & 0.28                         & 0.48                           & 0.42                           & 0.24                          & 0.47                            & 0.27                              \\
                      & CPT     & 6.89                         & 15.53                          & 33.01                          & 171.77                        & 201.16                          & 31.38                                 \\
&   \multicolumn{1}{r}{}        & \multicolumn{1}{r}{}        & \multicolumn{1}{r|}{}          & \multicolumn{1}{r}{}          & \multicolumn{1}{r}{}         & \multicolumn{1}{r|}{}           & \multicolumn{1}{r}{}               \\
\multirow{2}{*}{GALE} & RED     & 0.88                         & 0.31                           & 0.51                           & 0.17                          & 0.70                            & 0.27                          \\
                      & CPT     & 1.03                         & 2.27                           & 4.27                           & 8.27                          & 11.67                           & 31.38                             \\
  &   \multicolumn{1}{r}{}        & \multicolumn{1}{r}{}        & \multicolumn{1}{r|}{}          & \multicolumn{1}{r}{}          & \multicolumn{1}{r}{}         & \multicolumn{1}{r|}{}           & \multicolumn{1}{r}{}                 \\
\multirow{2}{*}{DCD}  & RED     & 0.14                         & 0.20                           & 0.26                           & 0.06                          & 0.21                            & 0.34                             \\
                      & CPT     & 3.27                         & 5.14                           & 7.98                           & 18.21                         & 19.02                           & 73.72                              \\
 &   \multicolumn{1}{r}{}        & \multicolumn{1}{r}{}        & \multicolumn{1}{r|}{}          & \multicolumn{1}{r}{}          & \multicolumn{1}{r}{}         & \multicolumn{1}{r|}{}           & \multicolumn{1}{r}{}                \\
\multirow{2}{*}{DCPL} & RED     & \textbf{0.13}                & \textbf{0.20}                  & \textbf{0.23}                  & \textbf{0.03}                 & \textbf{0.20}                   & \textbf{0.21}                      \\
                      & CPT     & \textbf{0.67}                & \textbf{0.97}                  & \textbf{3.89}                  & \textbf{4.08}                 & \textbf{6.85}                   & \textbf{23.34}                     \\
                      \hline
                      &         & \multicolumn{2}{c|}{$n=1,500$}                             & \multicolumn{3}{c|}{$n=3,000$}                                                                & \multicolumn{1}{c}{$n=5,000$}                                    \\
\cline{3-4}  \cline{5-7}       \cline{8-8}
\multirow{2}{*}{PSC}  & RED     & 0.17                         & 0.50                           & 0.30                           & 0.25                          & 0.24                            & 0.19                       \\
                      & CPT     & 5.31                         & 10.99                          & 18.95                          & 39.71                         & 42.22                           & 57.44                            \\
&   \multicolumn{1}{r}{}        & \multicolumn{1}{r}{}        & \multicolumn{1}{r|}{}          & \multicolumn{1}{r}{}          & \multicolumn{1}{r}{}         & \multicolumn{1}{r|}{}           & \multicolumn{1}{r}{}              \\
\multirow{2}{*}{PACE} & RED     & 0.13                         & 0.31                           & 0.25                           & 0.08                          & ---                             & 0.19                               \\
                      & CPT     & 8.29                         & 21.78                          & 40.44                          & 177.07                        & ---                             & 57.44                       \\
&   \multicolumn{1}{r}{}        & \multicolumn{1}{r}{}        & \multicolumn{1}{r|}{}          & \multicolumn{1}{r}{}          & \multicolumn{1}{r}{}         & \multicolumn{1}{r|}{}           & \multicolumn{1}{r}{}               \\
\multirow{2}{*}{GALE} & RED     & 0.15                         & 0.54                           & 0.34                           & 0.04                          & 0.35                            & 0.19                                     \\
                      & CPT     & 8.25                         & 8.28                           & 22.36                          & 21.85                         & 23.90                           & 57.44                                   \\
&   \multicolumn{1}{r}{}        & \multicolumn{1}{r}{}        & \multicolumn{1}{r|}{}          & \multicolumn{1}{r}{}          & \multicolumn{1}{r}{}         & \multicolumn{1}{r|}{}           & \multicolumn{1}{r}{}              \\
\multirow{2}{*}{DCD}  & RED     & 0.14                         & 0.17                           & 0.23                           & 0.04                          & 0.23                            & 0.28                            \\
                      & CPT     & 3.49                         & 5.64                           & 9.71                           & 17.35                         & 20.04                           & 78.23                                  \\
&   \multicolumn{1}{r}{}        & \multicolumn{1}{r}{}        & \multicolumn{1}{r|}{}          & \multicolumn{1}{r}{}          & \multicolumn{1}{r}{}         & \multicolumn{1}{r|}{}           & \multicolumn{1}{r}{}                \\
\multirow{2}{*}{DCPL} & RED     & \textbf{0.11}                & \textbf{0.18}                  & \textbf{0.19}                  & \textbf{0.03}                 & \textbf{0.19}                   & \textbf{0.16}                          \\
                      & CPT     & \textbf{3.35}                & \textbf{3.66}                  & \textbf{9.03}                  & \textbf{10.41}                & \textbf{15.72}                  & \textbf{23.63}                  \\
\hline
\end{tabular}}
\end{table}

Second, as the subsample size increases, the clustering accuracy of DCPL approaches that of the global CPL method. Notably, in the ca-HepPh network, the DCPL method exhibits superior performance compared to CPL. This remarkable improvement can be attributed to the distributed nature of the DCPL method, maximizing the pseudo-likelihood function across multiple subnetworks. It combines diverse local solutions from various workers, broadening the solution space to avoid local optima. Moreover, in terms of RED value, the PMOEA algorithm performs well in some datasets, while the DCPL method also shows comparable performance in these networks.

Third, regarding computational efficiency, the proposed DCPL method outperforms all other compared community detection algorithms. The computational advantage of the DCPL method can be attributed to two key factors: (1) based on the conditional pseudo-likelihood method, each worker conducts computationally feasible low-dimensional matrix operations when updating the local estimator; (2) the DCPL method easily obtains the global label estimator without complex label alignment in each iteration.

Despite being a model-based approach, the proposed method operates under specific network model assumptions, which is a common challenge encountered by other model-based algorithms \citep{lei2015consistency, cai2015robust, amini2013pseudo, yang2017stochastic, li2022hierarchical}. Additionally, the results of these experiments conclusively demonstrate the effectiveness of the proposed DCPL method for large-scale networks, offering both computational efficiency and high-quality community detection.


\subsection{Ablation Study}

We conduct an ablation study to evaluate the efficiency of each proposed component in our method. The proposed method consists of two parts: (a) the block-wise splitting approach and (b) the multi-round communication. The experimental results are presented in Table \ref{tab: ablation}.

\begin{table}[!htb]
\centering
\caption{An ablation study is conducted on the proposed block-wise splitting and multi-round communication. The clustering results are evaluated using the relative density (RED), the best RED under each subsample size setting for each network is highlighted in bold text.}\label{tab: ablation}
\resizebox{\columnwidth}{!}{
\begin{tabular}{cllccccc}
 \hline
\multirow{2}{*}{Network}     & \multirow{2}{*}{K} & \multicolumn{1}{c}{\multirow{2}{*}{communication}} & \multicolumn{2}{c}{$n=1,000$} &  & \multicolumn{2}{c}{$n=3,000$} \\
\cline{4-5} \cline{7-8}
                             &                    & \multicolumn{1}{c}{}                               & one-shot    & multi-round   &  & one-shot  & multi-round     \\
\multirow{2}{*}{ca-Heph}     & \multirow{2}{*}{6} & random splitting                                   & 1.072       & 0.355         &  & 0.787     & 0.225           \\
                             &                    & block-wise splitting                               & 0.355       & \textbf{0.220}         &  & 0.230     & \textbf{0.222}  \\
                             \\
\multirow{2}{*}{ca-AstroPh}  & \multirow{2}{*}{6} & random splitting                                    & 0.977       & 0.498         &  & 0.502     & 0.758           \\
                             &                    & block-wise-splitting                               & 0.206       & \textbf{0.156}         &  & 0.126     & \textbf{0.113}  \\
                             \\
\multirow{2}{*}{ca-CondMat}  & \multirow{2}{*}{6} & random splitting                                    & 0.801       & 0.508         &  & 0.631     & 0.370           \\
                             &                    & block-wise splitting                               & 0.418       & \textbf{0.273}         &  & 0.226     & \textbf{0.180}  \\
                             \\
\multirow{2}{*}{cit-HepPh}   & \multirow{2}{*}{9} & random splitting                                    & 0.701       & 0.375         &  & 0.491     & 0.310           \\
                             &                    & block-wise splitting                               & 0.056       & \textbf{0.031}         &  & 0.035     & \textbf{0.029}  \\
                             \\
\multirow{2}{*}{email-Enron} & \multirow{2}{*}{9} & random splitting                                   & 1.323       & 2.271         &  & 0.660     & 0.583           \\
                             &                    & block-wise splitting                               & 0.359       & \textbf{0.163}        &  & 0.188     & \textbf{0.137} \\
 \hline
\end{tabular}}
\end{table}

In Table \ref{tab: ablation}, the random splitting method refers to a modified version of the DCPL method, where the entire network is divided into subnetworks by randomly selecting several subsets of size $n$ and utilizing the connections within each subset. Conversely, the block-wise splitting method represents the original DCPL method that employs the proposed block-wise splitting approach during the dividing process. Regarding the communication in the distributed system, the one-shot method indicates a modified version of the DCPL method, where the communication between the master and workers occurs only once. In contrast, the multi-round method refers to the original DCPL method, where the master communicates with the workers in multiple rounds. In this ablation study, we set the maximum number of communications to be 10.

Table \ref{tab: ablation} presents the performance of eight combinations based on different network splitting methods, communication approaches, and subsample sizes in each worker. First, we compare the different network dividing approaches. The block-wise splitting method achieves the best RED for all datasets, indicating its superiority over the random splitting method in extracting connection information. Second, regarding the communication method, we observe that the clustering performance based on multi-round communication outperforms that of the one-shot communication for all datasets. Therefore, we recommend utilizing the multi-round communication approach in the proposed DCPL method to obtain improved community detection results. Third, as the subsample size $n$ increases, the RED of DCPL under each setting also increases. It is worth noting that the effect of the subsample size on the block-wise splitting method is less pronounced compared to the random splitting method. This demonstrates that the block-wise splitting method allows the DCPL method to have a more relaxed condition on the subsample size.

Based on these experiments, the proposed method demonstrates a significant improvement in the computational efficiency of large-scale community detection methods while maintaining desirable community detection accuracy.

\section{Concluding Remarks}

In this paper, we have developed computationally efficient distributed network community detection methods for large-scale networks. Namely, the DPL and DCPL methods for estimating the SBM and DCSBM, respectively. We have proposed a block-wise splitting method to effectively divide a large-scale network into several subnetworks. Consequently, the pseudo-likelihood or conditional pseudo-likelihood method can be applied to each subnetwork to obtain a local community label estimator. More importantly, the master can conveniently obtain a global estimator by gathering the local label estimators without alignment.

Furthermore, to ensure statistical accuracy, we have theoretically discussed the exact condition of the worker sample size, which could be as small as $O\{(\log{N})^{2}\}$. As a result, the computational complexity of the proposed method could be $O(N\log{N})$, which makes the analysis of large-scale networks more convenient. We have proved that the communication cost of the DPL is only $O(NR)$, which is of the same order as the recent communication-efficient distributed methods for independent samples, such as \cite{jordan2019communication}, \cite{fan2021communication}, and \cite{duan2022heterogeneity}. Finally, extensive numerical studies demonstrate an improvement in the computational efficiency of the proposed method.

We discuss two important directions for future research. First, we assume that the subnetworks within each worker have identical independent distributions. However, it can be generalized to allow for heterogeneous data distribution across workers. Second, our approaches can be extended to perform inference in models that have richer information rather than only community memberships, such as latent space models \citep{hoff2002latent, sewell2015latent} or the mixed membership stochastic block models \citep{airoldi2008mixed, jin2021optimal}.

\section{Acknowledgments}

This work was supported by the National Natural Science Foundation of China (grant numbers 12071477, 71873137, and 72271232) and the MOE Project of Key Research Institute of Humanities and Social Sciences (22JJD110001), and the fund for building world-class universities (disciplines) at Renmin University of China. The authors gratefully acknowledge the support of Public Computing Cloud, Renmin University of China.

\appendix

\section{Appendices}

The proof of Theorems \ref{the: sampleSize} is provided in Appendix A.1, and the proofs of Propositions \ref{pro: computational} and \ref{pro: communication} are provided in Appendices A.2 and A.3, respectively.

\subsection{Subsample Size of DPL}\label{app: subsample}

In this section, we provide the proof of Theorem \ref{the: sampleSize} by the following two steps. Under the assumptions in Theorem \ref{the: sampleSize}, we first prove that the in-worker node set $\mN_r$ covers $K$ blocks completely with probability at least $1-1/N$. Then, we demonstrate that the expected average degree of the subnetwork is $E(d_r)=\Omega(\log{N})$ with high probability.

{\sc Step 1.} We first represent event $\mN_r \in \mS_K$ by simple events. Specifically, we describe the event $X=\{\mN_r: \ \forall \ k \in [K], \ \exists \ i' \in \mN_r, s.t. \ z_{r,i'}=k\}$ by several simple events to calculate its probability. We denote $X_k = \{ \mN_r: \sum_{i' \in \mN_r} \bI(z_{r,i'}=k)>0\}$, for $k=1,\cdots, K.$ Then, we have $X= \bigcap_{k=1}^{K}X_{k}.$ We focus on calculating the probability of an event $X$ afeterward. Let $X^{c}$ denote the complement set for $X$. Subsequently, according to De Morgan's laws, $X^{c}= \bigcup_{k=1}^{K}X_k^{c}$. Therefore, based on the properties of the probability measure,
\beq\label{eq: prob}
P(X^{c})\le \sum_{k=1}^{K} P(X_k^{c}).
\eeq

Assume $N_{k}= \sum_{i=1}^{N}\bI(z_{i}=k)$ as the number of nodes in the $k$th cluster and define $N_{\rm min}=\min_{k} N_{k}$. Considering random simple sampling with replacement, the probability of choosing a node from the $k$-th block is $N_{k}/N$ for each sampling. Then, we have $P(X_{k}^{c})= (1-N_{k}/N)^{n}$, for $k=1,\cdots, K$. Thus, according to \eqref{eq: prob}, $P(X^{c}) \le \sum_{k=1}^{K}P(X_k^{c}) \le  K( 1- N_{{\rm min}}/N)^{n}$. In other words, $P(X) \ge 1- K (1- N_{{\rm min}}/N)^{n}$. Choose an integer value for $n$ such that $\epsilon\ge K (1- N_{{\rm min}}/N)^{n}$. Consequently, we have $n\ge \log(K/\epsilon)/\log{(1- N_{\rm min}/N)^{-1}}$. Based on Assumption \ref{ass: balance}, consider $\epsilon = 1/N$ and $K = O(1)$. Then, choose $n$ such that $n = \Omega(\log N)$. Consequently, we can conclude that $\mN_r \in \mS_K$ with a probability of at least $1-1/N$.

{\sc Step 2.} Consider that the network density is $\rho$ and under Assumptions \ref{ass: balance}-- \ref{ass: density}, we have $NE(d_r)=$ $E\big\{\sum_{i',j} a_{r, i'j}\big\}$ $= \Omega(n\rho)$. Furthermore, since $n=\Omega\{ (\log{N})/\rho\}$, we have $E(d)= \Omega(\log{N})$. Hence, we have proved Theorem \ref{the: sampleSize}.

\subsection{Computational Complexity of DPL}\label{app: computational}

Recall that in each iteration, the main computing task is accomplished by multiple workers in parallel. Specifically, the $(t+1)$th iteration comprises a two-step communication. Then, we analyze the computational time of each step as follows.

In the first step communication, each worker computes its local statistics, $\bO_{r}(\hat{\be}^{(s)})$ and $\bn_{r}(\hat{\be}^{(s)})$, which requires $O(Nn\rho)$ computational complexity. Next, the master calculates $(\hat{\bpi}^{(s)},\hat{\bLambda}^{(s)})$ and broadcasts to workers, which takes $O(R)$ running time. In the second step, each worker first applies the EM algorithm to estimate the parameters $(\bpi, \bLambda)$, which requires $O(Nn\rho)$ computation complexity. Subsequently, each worker updates the local estimates, requiring only $O(n)$ computational complexity. Lastly, the complexity of combing the local estimates in the master requires $O(N)$ complexity.

Thus, the total computational complexity of the DPL algorithm in each iteration is in the order of $O(Nn\rho)$. We have proved the Proposition \ref{pro: computational}

\subsection{Communication Cost of DPL}\label{app: communication}

First, we prove the statement regarding the communication cost of DPL in each iteration. Considering each iteration comprises a two-step communication, we analyze the communication cost of each procedure in the two-step communication. In the $(s+1)$th iteration:\\
(1) The master broadcasts the current label estimator $\hat{\be}^{(s)}$ to each worker, which costs $O(NR)$ bits for communication;\\
(2) Each worker calculates $\bO_{r}(\hat{\be}^{(s)})$ and $\bn_{r}(\hat{\be}^{(s)})$, and transmits the result to the master, which requires $O(R)$ bits;\\
(3) The master calculates $(\hat{\bpi}^{(s)},\hat{\bLambda}^{(s)})$ and broadcasts it to the workers which requires $O(R)$ bits;\\
(4) Each worker updates $\hat{e}^{(s+1)}_{r,i'}$, for all $1\le i' \le n$ and transmits $\hat{\be}^{(s+1)}_{r}$ to the master, which requires $O(N)$ bits.\\
Then, the master then updates the global estimator using $\hat{e}^{(s+1)}_{i}=\hat{e}^{(s+1)}_{r_i,w_i}$ for $1\le i \le N$. Consequently, the communication cost per iteration is $O(NR)$ bits. Hence, Proposition \ref{pro: communication} is proved.

\newpage
\bibliographystyle{asa}
\bibliography{reference}
\end{document}